\documentstyle[12pt]{article}
\newcommand{\be}{\begin{equation}}
\newcommand{\ee}{\end{equation}}
\newcommand{\beq}{\begin{eqnarray}}
\newcommand{\eeq}{\end{eqnarray}}
\newcommand{\lm}{\lambda}
\newcommand{\move}{\hspace{0.25in}}
\newcommand{\k}{\kappa}
\newcommand{\th}{\theta}
\newcommand{\om}{\omega}
\newcommand{\omz}{\omega_0}

\newcommand{\eps}{\epsilon}
\newcommand{\otwo}{{\rm O}(2)}
\newcommand{\sotwo}{{\rm SO}(2)}
\newcommand{\ord}[1]{{\rm \cal O}({#1})}
\newcommand{\lop}{{\rm \cal L}}
\newcommand{\alop}{{\rm \cal L}^\dagger}
\newcommand{\nop}[1]{{\rm \cal N}({#1})}

\newcommand{\g}{\gamma}
\newcommand{\gi}{\gamma^{-1}}
\newcommand{\flop}[1]{{\rm  L}_{#1}}
\newcommand{\aflop}[1]{{\rm  L}_{#1}^\dagger}
\newcommand{\aef}{\tilde{\Psi}}
\newcommand{\ef}{\Psi}
\newcommand{\eff}{\psi}
\newcommand{\aeff}{\tilde{\psi}}
\newcommand{\gef}{\Upsilon}
\newcommand{\geff}{\upsilon}
\newcommand{\agef}{\tilde{\Upsilon}}
\newcommand{\ageff}{\tilde{\upsilon}}
\newcommand{\e}{\eta}

\newcommand{\sfcn}[2]{\Lambda_{{#1}}\,({#2})}
\newcommand{\sfcnp}[2]{\Lambda'_{{#1}}\,({#2})}
\newcommand{\sfcnpp}[2]{\Lambda''_{{#1}}\,({#2})}
\newcommand{\sfcnppp}[2]{\Lambda'''_{{#1}}\,({#2})}
\newcommand{\pbnf}{Poincar\'e-Birkhoff normal form}
\newcommand{\otwoh}{{\rm O}(2)\times {\rm S}^1 }

\begin{document}
\baselineskip=14pt    \begin{titlepage}   \vspace*{0.0cm}

\begin{center}
{\bf Amplitude Expansions for Instabilities in Populations\\ of
Globally-Coupled Oscillators}\end{center}
\begin{center}
John David Crawford \\
Department of Physics and Astronomy\\
University of Pittsburgh\\
Pittsburgh, PA 15260
\end{center}
\vspace{0.5cm}

\centerline{\bf ABSTRACT}

\begin{quote}
We analyze the nonlinear dynamics near the incoherent state in a mean-field
model of coupled oscillators. The population is described by a Fokker-Planck
equation for the distribution of phases, and we apply center-manifold reduction
to obtain the amplitude equations for steady-state and Hopf bifurcation from
the equilibrium state with a uniform phase distribution. When the  population
is described by a native frequency distribution that is reflection-symmetric
about zero, the problem has circular symmetry. In the limit of zero extrinsic
noise, although the critical eigenvalues are embedded in the continuous
spectrum, the nonlinear coefficients in the amplitude equation remain finite in
contrast to the singular behavior found in similar instabilities described by
the Vlasov-Poisson equation. For a bimodal reflection-symmetric distribution,
both types of bifurcation are possible and they coincide at a codimension-two
Takens Bogdanov point. The steady-state bifurcation may be supercritical or
subcritical and produces a time-independent synchronized state. The Hopf
bifurcation produces both supercritical stable standing waves and supercritical
unstable travelling waves. Previous work on the Hopf bifurcation in a bimodal
population by  Bonilla, Neu, and Spigler and Okuda and Kuramoto predicted
stable travelling waves and stable standing waves, respectively. A comparison
to these previous calculations shows that the prediction of stable travelling
waves results from a failure to include all unstable modes.
 \end{quote}

\begin{center}
keywords: oscillators, bifurcation, symmetry
\end{center}
\vfill
\begin{center}
October 28, 1993
\end{center}
\end{titlepage}
\tableofcontents
\baselineskip=24pt

\section{Introduction}
\move The dynamics of a collection of weakly coupled limit cycle oscillators
can be analyzed by treating the oscillator phases
$\{\theta_1,\ldots,\theta_N\}$ as an autonomous dynamical system. In the
context of biological oscillators this approach was advocated by Winfree and
subsequently developed by Kuramoto for
reaction-diffusion systems.\cite{win,kur}  In particular, Kuramoto formulated
the widely studied model for phase dynamics:
\be
\dot{\theta}_i=\omega_i+\frac{K}{N}\sum^{N}_{j=1} \sin(\theta_j-\theta_i) +
\xi_i(t)\label{eq:kur}
\ee
where the natural frequencies $\{\omega_i\}$ are randomly distributed according
to a density $g(\omega)$. The effect of adding noise $\xi_i(t)$ to each phase
was considered later by Sakaguchi.\cite{sak}

For large $N$, this model describes a transition to collective synchronized
behavior as $K$ increases above a critical threshold $K_c$. More precisely,
in terms of the order parameter,
\be
R(t)\,e^{i\psi(t)}=\frac{1}{N}\sum^{N}_{j=1}\;e^{i\theta_j (t)},\label{eq:op}
\ee
one finds numerically a transition from an ``incoherent'' state with
$R(t)\approx 0$ to a partially synchronized state with $R(t)\sim\sqrt{K-K_c}$
for $K>K_c$. For $g(\omega)$ chosen to be unimodal and symmetric about
$\omega=0$, Kuramoto obtained an analytic expression for the order parameter
$r$ as a function of the coupling $K$. His result agrees with numerical
simulations and yields a simple expression for the critical coupling in the
absence of noise
\be
K_c=\frac{2}{\pi g(0)}.\label{eq:kcrit}
\ee
This transition has been analyzed by several authors\cite{kur84}-\cite{bns} and
additional references can be found in Kuramoto and Nishikawa.\cite{kn} If
$g(\omega)$ has compact support, then for sufficiently large $K$ there can be a
transition to a completely synchronized state with $R\approx 1$.\cite{erm,hw}

Since the incoherent state and the partially synchronized state emerge when the
initial value problem is solved numerically, each must be stable in the
appropriate range of $K$ at least in an operational sense.
However the theoretical explanation of this stability has proved to be rather
subtle even for the incoherent state, but there has been significant recent
progress in the work of Strogatz and Mirollo.\cite{sm,smm}  Following
Sakaguchi, they considered the large $N$ limit of (\ref{eq:kur}) and studied
the Fokker-Planck equation
\be
\frac{\partial\rho}{\partial t}+\frac{\partial(\rho v)}{\partial \theta}=D
\frac{\partial^2\rho}{\partial \theta^2}\label{eq:eveqn}
\ee
for the distribution of oscillators. The density $\rho(\theta,\omega,t)$ is
defined so that $N\,g(\omega)\rho(\theta,\omega,t)\,d\theta\,d\omega$
describes the number of oscillators with natural frequencies in
$[\omega,\omega+d\omega]$ and phases in $[\theta,\theta+d\theta]$. Thus
$\rho(\theta,\omega,t)\,d\theta$ denotes the fraction of oscillators with
natural frequency $\omega$ and phase in $[\theta,\theta+d\theta]$ and satisfies
the normalization
\be
\int^{2\pi}_0\,d\th\rho(\th,\om,t)=1.\label{eq:norm}
\ee
The identity
\be
R(t)\sin(\psi-\theta_i)=\frac{1}{N}\sum^{N}_{j=1}\;\sin(\theta_j -\theta_i)
\ee
allows the phase velocity (\ref{eq:kur}) of an oscillator to be written as
\be
v(\th,\om,t)=\om +K R(t)\sin(\psi-\th),\label{eq:vel}
\ee
and the order parameter (\ref{eq:op}) can be re-expressed in terms of the
density
\be
R(t)e^{i\psi(t)}= \int^{2\pi}_0\,d\th\int^{\infty}_{-\infty}\,d\om
\rho(\th,\om,t)\,g(\om)\,e^{i\th}.\label{eq:ord}
\ee
The diffusion coefficient in (\ref{eq:eveqn}) reflects the Gaussian noise terms
$\xi_i(t)$ defined by
\beq
<\xi_i(t)>&=&0\\
<\xi_i(s)\,\xi_j(t)>&=&2D\delta_{ij}\delta(s-t).
\eeq

Equations (\ref{eq:eveqn}) and (\ref{eq:vel}) - (\ref{eq:ord}) provide a
continuum description of the oscillator population for which issues of
stability and bifurcation can be analyzed in some detail. The incoherent state
is described by the uniform distribution
$\rho_0={1}/{2\pi}$
and defines an equilibrium for (\ref{eq:eveqn}) since $R=0$ at $\rho_0$. By
defining $\eta$, the deviation from $\rho_0$,
$\rho(\th,\om,t)=\rho_0 +\eta(\th,\om,t)$
and eliminating $v$ in (\ref{eq:eveqn}), the model can be rewritten as a single
equation for $\eta$:
\be
\frac{\partial\eta}{\partial t}=\lop\eta +\nop{\eta}\label{eq:dyn}
\ee
where
\be
\lop\eta\equiv D \frac{\partial^2\eta}{\partial \theta^2} -
\om\frac{\partial\eta}{\partial \theta} + \frac{K\rho_0}{2}
\left[e^{-i\theta}\int^{2\pi}_0\,d\th'\int^{\infty}_{-\infty}\,d\om'
\eta(\th',\om',t)\,g(\om')\,e^{i\th'} + cc\right] \label{eq:lop}
\ee
and the nonlinear terms are
\beq
\nop\e&=&\frac{K}{2}
\left\{e^{i\theta}\left(\e - i\frac{\partial\eta}{\partial \theta}\right)
\int^{2\pi}_0\,d\th'\int^{\infty}_{-\infty}\,d\om'
\eta(\th',\om',t)\,g(\om')\,e^{-i\th'} \nonumber\right.\\
&&\left.\hspace{0.3in}
+e^{-i\theta}\left(\e + i\frac{\partial\eta}{\partial \theta}\right)
\int^{2\pi}_0\,d\th'\int^{\infty}_{-\infty}\,d\om'
\eta(\th',\om',t)\,g(\om')\,e^{i\th'}\right\}.
\label{eq:nop}
\eeq
Note that the normalization (\ref{eq:norm}) of $\rho$ implies
\be
\int^{2\pi}_0\,d\th\,\eta(\th,\om,t)=0.\label{eq:norm1}
\ee

An important qualitative feature of this evolution equation is its symmetry.
The group $\otwo$ is generated by rotations
\be
\phi\cdot(\th,\om)=(\th +\phi,\om)\label{eq:phi}
\ee
and reflections
\be
\k\cdot(\th,\om)=-(\th,\om)\label{eq:kap}
\ee
which act on functions $\eta(\th,\om)$ in the usual way:
\be
(\g\cdot\eta)(\th,\om)=\eta(\gi\cdot(\th,\om))\hspace{0.5in}\g\in\otwo.
\label{eq:rep}
\ee
Provided $g(\om)$ satisfies
\be
g(\om)=g(-\om),\label{eq:g}
\ee
the equation for $\eta$ (\ref{eq:dyn}) has $\otwo$ symmetry; more precisely
$\g\cdot(\lop\eta)=\lop(\g\cdot\eta)$ and $\g\cdot\nop{\eta}=\nop{\g\cdot\eta}$
for all $\g\in\otwo$. A realistic population of oscillators will not have zero
mean frequency as implied by (\ref{eq:g}), but if their frequency distribution
$G(\om_{\rm lab})$ is symmetric about the mean $\overline{\om}$ then in the
rotating frame defined by
\be
(\theta,\om)=(\theta_{\rm lab}-\overline{\om}\,t,\,\om_{\rm
lab}-\overline{\om}) \label{eq:rotf}
\ee
the frequency distribution $g(\om)=G(\om+\overline{\om})$ will satisfy
(\ref{eq:g}). The densities in the lab frame and in the rotating frame are
related by\footnote{In the lab frame the evolution equation for $\rho_{\rm
lab}(\theta_{\rm lab},\,\om_{\rm lab},t)$ has rotation symmetry ${{\rm SO}(2)}$
but not reflection symmetry. Calculations done using lab frame variables have
puzzling features that are non-generic for ${{\rm SO}(2)}$ symmetry and have
their origin in the full $\otwo$ symmetry of the rotating frame.}
\be
\rho(\theta,\,\om,t)=
\rho_{\rm lab}(\theta+\overline{\om}\,t,\,\om+\overline{\om},t).
\label{eq:lab}
\ee
If $g$ lacks reflectional invariance (\ref{eq:g}) then the symmetry in the
rotating frame is reduced to the rotations ${{\rm SO}(2)}$.\cite{kno}

Strogatz and Mirollo provide a linear stability analysis of $\rho_0$ focussing
especially on populations with even distributions (\ref{eq:g}) that decrease
monotonically away from $\omega=0$.\cite{sm} They show that $\lop$ has a
continuous spectrum with real part equal to $-D$, and that it may also have
point spectrum (eigenvalues) depending on the coupling strength. For $K$
sufficiently small, the point spectrum is empty, but as the coupling increases
a real eigenvalue $\lambda$ emerges from the continuous spectrum and for
$K>K_c$ moves into the right half plane $\lambda>0$ signifying linear
instability of $\rho_0$. The condition $\lambda=0$ determines a critical
coupling
\be
K_c=2\left[\int^{\infty}_{-\infty}\, \frac{D\,g(\om)\,d\om}
{D^2+\om^2} \right]^{-1}
\ee
that  agrees with results for $K_c$ derived by Sakaguchi using different
arguments and reduces to (\ref{eq:kcrit}) when $D\rightarrow0$. Thus the onset
of synchronization is shown to coincide with the appearance of a linearly
unstable mode for the Fokker-Planck equilibrium. Subsequently Strogatz et al.
demonstrated that for $D=0$ and $K<K_c$, the linearized equation predicts
$\rho_0$ is stable in the sense that $R(t)$ decays via a ``phase mixing''
mechanism that is qualitatively similar to Landau damping in a collisionless
plasma.\cite{smm}

Recently Bonilla, Neu, and Spigler investigated bifurcations from $\rho_0$
assuming a reflection-symmetric frequency distribution (\ref{eq:g}) but
relaxing the requirement that $g(\om)$ have a single peak.\cite{bns} For
bimodal distributions they showed that $\lop$ could have complex eigenvalues so
that the instability of $\rho_0$ occurs via Hopf bifurcation. Amplitude
equations were derived to describe the nonlinear development of the
instabilities, but unfortunately the Hopf analysis retains only half of the
unstable eigenvectors and as a result nonlinear travelling and standing wave
solutions were overlooked. The fact that the $\otwo$ symmetry leads to
degenerate eigenvalues of multiplicity-two was apparently not recognized.

Independently, Okuda and Kuramoto analyzed the onset of collective behavior in
a population comprised of two sub-populations using a different model; each
sub-population had a unique frequency in the absence of coupling and was
described by a number density.\cite{ok} The instability of the incoherent state
was studied under several assumptions on the couplings between the oscillators.
In the regime where the coupling between the populations equals the coupling
within a population, their model is a special case of (\ref{eq:eveqn}). They
also find that instability can arise from either real or complex eigenvalues
and derived amplitude equations for the nonlinear behavior. Their equations
include both travelling and standing wave solutions although the role of
symmetry in the problem is not discussed.

In this paper I re-examine the bifurcation problem and derive amplitude
equations for both steady state and Hopf bifurcation using center manifold
reduction. In Section II, the spectrum of $\lop$ is discussed to clarify the
treatment of complex eigenvalues and to discuss the occurrence of
non-semi-simple real eigenvalues. These arise naturally at the multi-critical
point where the Hopf and steady-state bifurcations coalesce. This
codimension-two point corresponds to a so-called Takens-Bogdanov bifurcation
and the relevant amplitude equations will be investigated in future work. The
discussion of Section II is facilitated by the similarities of $\lop$ to the
linear operator for the Vlasov-Poisson equation which has been discussed
elsewhere.\cite{ch}

In Section III, the amplitude equations are derived and their limiting form as
$D\rightarrow0$ determined. The results are strikingly different from analogous
amplitude equations for Vlasov instabilities; in the collisionless plasma case
the coefficients diverge as the critical eigenvalues approach the continuum but
in this model the coefficients have finite limits.\cite{jdc}
In the nonlinear theory, the presence of $\otwo$ symmetry is exploited to take
advantage of the well-established theory of bifurcation in systems with
circular symmetry.\cite{gss,craw} For a reflection-symmetric bimodal population
comprised of two Lorentzians, we evaluate the cubic normal form coefficients.
For the steady-state bifurcation the appearance of the synchronized state can
be either supercritical or subcritical. In the Hopf bifurcation we find that
the standing waves are supercritical and stable while the travelling waves are
supercritical and unstable. These conclusions are compared to previous
studies.\cite{ok,bns}

\section{Linear theory}
\move The Fourier expansion of $\e$,
\be
\e(\th,\om)=\sum_{l=-\infty}^{\infty}\,\e_l(\om)\,e^{il\th}\label{eq:fexp}
\ee
allows the linear operator (\ref{eq:lop}) to be re-expressed as
\be
\lop\;\e= \sum_{l=-\infty}^{\infty}\,(\flop l\,\e_l)\,e^{il\th}\label{eq:flp1}
\ee
where
\be
\flop l\;\e_l=-il\,(\om-ilD)\,\e_l + \pi\rho_0 K\,(\delta_{l,1}+\delta_{l,-1})
\int^{\infty}_{-\infty}\,d\om'g(\om')\,\e_l(\om').\label{eq:flp2}
\ee
Note that the normalization condition (\ref{eq:norm1}) on $\e$  implies
\be
\e_{l=0}\equiv0.\label{eq:mean}
\ee

An adjoint operator $(A,\lop  B)\equiv(\alop A, B)$
can be defined for the inner product
\be
(A,B)\equiv\int^{2\pi}_{0}\,d\th
\int^{\infty}_{-\infty}\,d\om\,A(\th,\om)^\ast\,B(\th,\om).\label{eq:inner}
\ee
In terms of the Fourier expansion  $A=\sum A_l e^{il\theta}$, this definition
yields
\be
\alop A= \sum_{l=-\infty}^{\infty}\,(\aflop l\,A_l)\,e^{il\th}\label{eq:aflp1}
\ee
where
\be
(\aflop l A_l)(\om)= il\,(\om+ilD)\,A_l(\om) + \frac{g(\om)K}{2}
(\delta_{l,1}+\delta_{l,-1})
\,\int^{\infty}_{-\infty}\,d\om'\,A_l(\om').\label{eq:aflp}
\ee
It is convenient to introduce a concise notation for the integration over
$\omega$ in (\ref{eq:inner})
\be
<A,B>\equiv\int^{\infty}_{-\infty}\,d\om\,A^\ast\,B;\label{eq:innerw}
\ee
then $\flop l$ and $\aflop l$ satisfy $<A,\flop l B>=<\aflop lA, B>.$

The eigenvalue problem for $\lop$,
\be
\lop\,\ef=\lm\ef,\label{eq:eig1}
\ee
can be treated on each Fourier subspace separately so we set
\be
\ef(\th,\om)=e^{il\th}\,\eff(\om),\label{eq:ef}
\ee
and reduce (\ref{eq:eig1}) to
\be
\flop l\,\eff=\lm\eff.
\label{eq:eig2}
\ee
After scaling the eigenvalue by
$\lm=-il\,z,$
the eigenvalue equation (\ref{eq:eig2}) becomes
\be
[(\om-z)-ilD]\,\eff=\frac{-ilK}{2}(\delta_{l,1}+\delta_{l,-1})
\int^{\infty}_{-\infty}\,d\om'g(\om')\,\eff(\om').\label{eq:eigex0}
\ee

The first observation is that $|l|=1$ is a necessary condition for $\psi$ to be
a well-behaved eigenfunction.\cite{sm} When $|l|\neq1$ then (\ref{eq:eigex0})
becomes $[(\om-z)-ilD]\,\eff=0,$
and we have $\eff\equiv0$ unless ${\rm Im}\,z=-lD$. For ${\rm Im}\,z=-lD$ then
$[(\om-z)-ilD]$ is non-zero except at $\om={\rm Re}\,z$, so the only solutions
are delta functions $\delta\,(\om-{\rm Re}\,z).$
These solutions correspond to a continuous spectrum for $\flop l$ along
Re$\,\lm=l\;{\rm Im}\,z=-l^2D$; as $D\rightarrow0$ the continuous spectrum
moves onto the imaginary axis. For $|l|\neq1$ this is the only contribution to
the spectrum of $\lop$ and it does not depend on the coupling $K$ or on the
frequency distribution $g(\om)$.

For $|l|=1$ it is possible to have eigenfunctions that are non-singular; their
occurrence does depend on $K$ and $g(\om)$. Now the eigenvalue equation
(\ref{eq:eigex0}) becomes
\be
[(\om-z)-ilD]\,\eff=\frac{-ilK}{2}
\int^{\infty}_{-\infty}\,d\om'g(\om')\,\eff(\om').\hspace{0.2
in}\;\;\;\;(|l|=1) \label{eq:eigl1}
\ee
When ${\rm Im}\,z= -lD$, then as before we get singular solutions corresponding
to a continuous spectrum along ${\rm Re}\,\lm=-D$; the detailed form of these
solutions will not be needed. If Im$\,z\neq -lD$ then $(\om-z)-ilD$ cannot
vanish, and we have
\be
\eff=\frac{-ilK/2}{(\om-z)-ilD}
\int^{\infty}_{-\infty}\,d\om'g(\om')\,\eff(\om').\label{eq:efcn0}
\ee
If the integral vanishes then $\eff\equiv0$ so we may assume the integral is
non-zero and normalize $\eff$ such that
\be
\int^{\infty}_{-\infty}\,d\om'g(\om')\,\eff(\om')=1.\label{eq:efnm}
\ee
Consistency between (\ref{eq:efcn0}) and (\ref{eq:efnm}) requires $z$ to be a
root of the ``spectral function''
\be
\sfcn l{z}\equiv 1+\frac{ilK}{2}
\int^{\infty}_{-\infty}\,d\om'\frac{g(\om')}{\om' -z-ilD}. \;\;\;\;\hspace{0.2
in}(|l|=1).\label{eq:Lam}
\ee

The spectral function $\Lambda_l$ satisfies two identities; for arbitrary
$g(\om)$ we have
\be
\sfcn l{z}^\ast=\sfcn {-l}{z^\ast}\label{eq:lam1}
\ee
so when $\sfcn l{z_0}=0$ determines an eigenvalue $\lm =-ilz_0$ then $\sfcn
{-l}{z_0^\ast}=0$ determines the complex-conjugate eigenvalue $\lm^\ast$. In
addition when the frequency distribution is reflection-symmetric (\ref{eq:g}),
the problem has $\otwo$ symmetry and $\Lambda_l$ satisfies
\be
\sfcn l{z}^\ast=\sfcn {l}{-z^\ast}.\label{eq:lam2}
\ee
Now roots come in pairs whenever ${\rm Re}\,z\neq0$, i.e. whenever $\lambda$ is
complex.

The roots of $\sfcn l{z}$ determine the eigenvalues and the nature of the root
determines whether the eigenvalue is semisimple or not. Recall that the
projection operator $P_\lambda$ onto the generalized eigenspace $E_\lambda$ for
an isolated eigenvalue $\lambda=-ilz_0$ is defined by a contour
integral,\cite{simon}
\be
P_\lambda A=\frac{1}{2\pi i}\int_{\Gamma_0}\,d\lambda'\,(\lm'-\lop )^{-1}A;
\ee
where $\Gamma_0$ is a small loop enclosing $\lambda$ in a counter-clockwise
sense. The resolvent operator $(\lm-\lop )^{-1}$ can be expressed
as\footnote{In this sum, the term $l=0$ may be omitted since the relevant space
of functions satisfies (\ref{eq:mean}).}
\be
(\lm-\lop )^{-1}A= \sum_{l=-\infty}^{\infty}\,(R_l(\lm)\,A_l)\,e^{il\th}
\label{eq:ressum}
\ee
where $R_l(\lm)\equiv(\lm-\flop l)^{-1}$
denotes the resolvent for $\flop l$ and can be calculated explicitly
\be
R_l(\lm)A_l=
\frac{A_l(\om)}{il\om +l^2D+\lm}+\frac{(\delta_{l,1}+\delta_{l,-1})K/2}
{\sfcn{l}{il\lm}(il\om +l^2D+\lm)}
\int^{\infty}_{-\infty}\,d\om'\,\frac{g(\om')\,A_l(\om')}{il\om'
+l^2D+\lm}.\label{eq:res}
\ee
With (\ref{eq:ressum}) $P_\lm$ may be simplified to
\be
P_\lambda A=\frac{e^{i\th}}{2\pi i}\int_{\Gamma_0}\,d\lambda'\,R_1(\lm')\,A_1+
\frac{e^{-i\th}}{2\pi i}\int_{\Gamma_0}\,d\lambda'\,R_{-1}(\lm')\,A_{-1}
\label{eq:proj}
\ee
since eigenvalues can only occur for $|l|=1$.
If $E_\lambda$, the range of $P_\lambda$, is one-dimensional then $\lambda$ is
simple; when the dimension is greater than one and the eigenvectors of $\lop$
span $E_\lambda$ the eigenvalue is semisimple. If dim$E_\lambda>1$, and the
eigenvectors do not span $E_\lambda$, then $\lambda$ is non-semisimple.
Simple roots of $\sfcn l{z}$ lead to simple or semisimple eigenvalues and
multiple roots give eigenvalues that are non-semisimple.  We discuss this point
in detail for the case of a reflection-symmetric frequency distribution and
$\otwo$ symmetry.

\subsection{Simple roots and semisimple eigenvalues}

\move For a simple root $z_0$ of  $\sfcn 1{z}$, we have
\be
\sfcn 1{z_0}=0\hspace{1.0in}\sfcnp{1}{z_0}\neq0\label{eq:r1}\\
\ee
where $\Lambda'$ denotes the derivative of $\sfcn 1{z_0}$ with respect to $z$.
The identities (\ref{eq:lam1}) - (\ref{eq:lam2}) imply the additional simple
roots
\be
\sfcn 1{-z_0^\ast}=0\hspace{1.0in}
\sfcn{-1}{z_0^\ast}=0\hspace{1.0in}\sfcn{-1}{-z_0}=0.
\ee
If ${\rm Re}\,z_0=0$ ( a real eigenvalue) then $z_0$ and $-z_0^\ast$ are the
same root and roots $(z_0^\ast,-z_0)$ also coincide. When ${\rm Re}\,z_0\neq0$
(complex eigenvalues) then these roots are all distinct.

\subsubsection{Real eigenvalues (${\rm Re}\,z_0=0$)}

\move With ${\rm Re}\,z_0=0$, the roots $z_0=$  and $z_0^\ast$ for $l=\pm1$
lead to linearly independent eigenvectors for the real eigenvalue
$\lambda=-iz_0$. From (\ref{eq:ef}) and (\ref{eq:efcn0}), these vectors are
\be
\ef(\th,\om)=e^{i\th}\,\eff(\om),\label{eq:efa}
\ee
where
\be
\eff(\om)=\frac{-iK/2}{(\om-z_0)-iD}\label{eq:efcna}
\ee
and the conjugate function
\be
\ef(\th,\om)^\ast=e^{-i\th}\,\eff(\om)^\ast.\label{eq:efacc}
\ee
Note, that these eigenvectors are related by reflection
$\ef(\th,\om)^\ast=(\k\cdot\ef)(\th,\om)$.
Corresponding to $\ef$ and $\k\cdot\ef$, there are adjoint eigenvectors
\beq
\alop\aef&=&\lambda\aef\label{eq:aeveqn}\\
\alop(\k\cdot\aef)&=&\lambda(\k\cdot\aef)\label{eq:aeveqn2}
\eeq
where
\be
\aef(\th,\om)=\frac{e^{i\th}}{2\pi}\aeff(\om)\label{eq:aef}
\ee
with
\be
\aeff(\om)=\frac{-g(\om)}{\sfcnp{1}{z_0}^\ast(\om-z_0^\ast+iD)}.\label{eq:aeff}
\ee
These eigenvectors are conjugate $\k\cdot\aef=\aef^\ast$ and satisfy
\beq
(\aef,\ef)&=&(\k\cdot\aef,\k\cdot\ef)=1\label{eq:orthog1}\\
(\aef,\k\cdot\ef)&=&(\k\cdot\aef,\ef)=0.\label{eq:orthog2}
\eeq
For example, the eigenvalue equation (\ref{eq:aeveqn}) can be verified directly
from the definitions (\ref{eq:aflp1}) - (\ref{eq:aflp}):
\beq
\alop\aef&=&\frac{e^{i\th}}{2\pi}\,\aflop 1\aeff(\om)\nonumber\\
&=&\frac{e^{i\th}}{2\pi\sfcnp{1}{z_0}^\ast}
\left[-\frac{ig(\om)}{(\om-z_0^\ast+iD)}\right.\nonumber\\
&&\left.\hspace{0.5in}-\frac{g(\om)
K}{2}\int^{\infty}_{-\infty}\,d\om'\frac{g(\om')}{(\om'-z_0^\ast+iD)}
\right]\nonumber\\
&=&\frac{e^{i\th}}{2\pi\sfcnp{1}{z_0}^\ast} \left[ -ig(\om)\right.\nonumber\\
&&\left.\hspace{0.5in}-\frac{ig(\om)z_0^\ast}{(\om-z_0^\ast+iD)}
-ig(\om)(\sfcn{-1}{z_0^\ast}-1)\right]\nonumber\\
&=&iz_0^\ast\,\frac{e^{i\th}}{2\pi\sfcnp{1}{z_0}^\ast}
\left(\frac{-g(\om)z_0^\ast}{(\om-z_0^\ast+iD)}\right)\nonumber\\
&=&\lambda\aef
\eeq
since $\lambda=\lambda^\ast=-iz_0.$

Since $\sfcn 1{z_0}=0$ corresponds to a simple pole, the result of projecting
$\eta$ onto $\ef$ is easily evaluated. From (\ref{eq:res}) - (\ref{eq:proj}) we
have
\beq
P_\lambda\;\eta&=& \frac{e^{i\th}}{2\pi
i}\int_{\Gamma_0}\,d\lambda'\,R_1(\lm')\eta_1 +  \frac{e^{-i\th}}{2\pi
i}\int_{\Gamma_0}\,d\lambda'\,R_{-1}(\lm')\eta_{-1}\nonumber\\
&&\nonumber\\
&=& \frac{e^{i\th}}{2\pi i}\int_{\Gamma_0}\,d\lambda' \frac{K/2}
{\sfcn{1}{i\lm'}(i\om +D+\lm')}
\int^{\infty}_{-\infty}\,d\om'\,\frac{g(\om')\,\eta_1(\om')}{i\om'+D+\lm'}
\nonumber\\
&&+\frac{e^{-i\th}}{2\pi i}\int_{\Gamma_0}\,d\lambda' \frac{K/2}
{\sfcn{-1}{-i\lm'}(-i\om +D+\lm')} \int^{\infty}_{-\infty}\,d\om'\,
\frac{g(\om')\,\eta_{-1}(\om')}{-i\om'+D+\lm'}\nonumber\\
&&\nonumber\\
&=&e^{i\th}\frac{K/2}{i\sfcnp{1}{z_0}(i\om +D-iz_0)}
\int^{\infty}_{-\infty}\,d\om'\,\frac{g(\om')\,\eta_1(\om')}{i\om'+D-iz_0}
\nonumber\\
&&+e^{-i\th}\frac{K/2}{-i\sfcnp{-1}{z_0^\ast}(-i\om +D+iz_0^\ast)}
\int^{\infty}_{-\infty}\,d\om'\,
\frac{g(\om')\,\eta_{-1}(\om')}{-i\om'+D+iz_0^\ast}\nonumber\\
&&\nonumber\\
&=&e^{i\th}\eff(\om)\,<\aeff,\eta_1>+
e^{-i\th}\eff(\om)^\ast\,<\aeff^\ast,\eta_{-1}>\nonumber\\
&=&(\aef,\eta)\,\ef(\th,\om) +(\k\cdot\aef,\eta)\,\k\cdot\ef(\th,\om).
\label{eq:ssev}
\eeq
The eigenvectors $\ef$ and $\k\cdot\ef$ clearly span the range of $P_\lm$.

\subsubsection{Complex eigenvalues (${\rm Re}\,z_0\neq0$)}

\move When ${\rm Re}\,z_0\neq0$, then $z_0$ and $-z_0$ are distinct roots of
$\sfcn 1{z}$ and $\sfcn {-1}{z}$, respectively, that determine two linearly
independent eigenvectors for the complex eigenvalue $\lambda=-iz_0$. These
eigenvectors are $\ef$ given by (\ref{eq:efa}) - (\ref{eq:efcna}) and the
reflected function
\be
\k\cdot\ef=e^{-i\th}\eff(-\om).\label{eq:hopfev2}
\ee
This second eigenvector was omitted in the Hopf analysis of Bonilla {\em et
al.}\cite{bns} The remaining roots $-z_0^\ast$ and $z_0^\ast$ for $\Lambda_1$
and $\Lambda_{-1}$ determine the conjugate eigenvectors $\ef^\ast$ and
$(\k\cdot\ef)^\ast$ corresponding to the conjugate eigenvalue
$\lambda^\ast=iz_0^\ast$.
The adjoint eigenvectors $\aef$ and $\k\cdot\aef$ have the same form
(\ref{eq:aef}) - (\ref{eq:aeff}), however now (\ref{eq:aeveqn}) -
(\ref{eq:aeveqn2}) become
\beq
\alop\aef&=&\lambda^\ast\aef\label{eq:aeveqn3}\\
\alop(\k\cdot\aef)&=&\lambda^\ast(\k\cdot\aef);\label{eq:aeveqn4}
\eeq
the adjoint eigenvectors for $\lambda$ are the conjugate functions $\aef^\ast$
and $(\k\cdot\aef)^\ast$. The orthogonality relations (\ref{eq:orthog1}) -
(\ref{eq:orthog2}) remain valid.

The projection $P_\lm\,\eta$ can be evaluated as before (\ref{eq:ssev}) except
that now the relevant root (pole) for $l=-1$ is $\sfcn {-1}{-z_0}=0$. The final
result has the same form
\be
P_\lm\,\eta= (\aef,\eta)\,\ef +(\k\cdot\aef,\eta)\,\k\cdot\ef\label{eq:hev1}
\ee
A similar evaluation of $P_{\lm^\ast}\,\eta$ gives
\be
P_{\lm^\ast}\,\eta=(P_{\lm}\,\eta)^\ast.\label{eq:hev2}
\ee
In each case the eigenvalue is semisimple with multiplicity-two.

\subsection{Multiple roots and non-semisimple eigenvalues}

\move A single root, $\sfcn {1}{z_0}=0$, determines exactly one eigenvector
(\ref{eq:efa}) - (\ref{eq:efcna}) for the eigenvalue $\lm=-iz_0$. There can be
additional linearly independent eigenvectors for the same eigenvalue $\lm$ only
if $\Lambda_{-1}$ simultaneously has a root at $-z_0$. In any event, a given
eigenvalue can never have more than two eigenvectors. However if the root $z_0$
of $\Lambda_{1}$ is not simple, the range of $P_\lm$ turns out to have
dimension greater than the number of eigenvectors because there are generalized
eigenvectors in addition to the true eigenvectors. A generalized eigenvector
$\gef$ satisfies
\beq
(\lop-\lm)^j\gef&=&0\label{eq:gef1}\\
(\lop-\lm)^{j-1}\gef&\neq&0\label{eq:gef2}
\eeq
for some integer $j\geq2$. Note that (\ref{eq:gef1}) simply says that
$(\lop-\lm)^{j-1}\gef$ is an eigenvector for $\lm$.

Since degenerate roots of $\Lambda_{1}$ arise when simple roots collide, we
expect generalized eigenvectors to enter the linear problem whenever
eigenvalues collide as parameters are varied. The connection between degenerate
roots and generalized eigenvalues can be made explicit as follows. Consider the
simplest situation $j=2$ and let $\ef=e^{i\th}\eff$ be a true eigenvector
corresponding to $\sfcn {1}{z_0}=0$. We seek a generalized eigenvector
$\gef=e^{i\th}\geff$ satisfying
\beq
(\lop-\lm)\gef&=&\ef\label{eq:gef3}\\
(\lop-\lm)^2\gef&=&0;\label{eq:gef4}
\eeq
note that (\ref{eq:gef3}) implies (\ref{eq:gef4}). From the definition of
$\lop$, (\ref{eq:gef3}) can be rewritten as
\be
(\om-z_0-iD)\geff +
\frac{iK}{2}\int^{\infty}_{-\infty}\,d\om'g(\om')\,\geff(\om') =
\frac{K/2}{\om-z_0-iD};\label{eq:gef5}
\ee
upon multiplying  by $g(\om)/(\om-z_0-iD)$ and integrating (\ref{eq:gef5})
becomes
\be
\sfcn {1}{z_0} \int^{\infty}_{-\infty}\,d\om'g(\om')\,\geff(\om')=-i\sfcnp
{1}{z_0}.\label{eq:gef6}
\ee
Since $\sfcn {1}{z_0}=0$, the assumption (\ref{eq:gef3}) that a generalized
eigenvector $\gef$ exists leads immediately to a contradiction if the root is
simple, i.e. if $\sfcnp {1}{z_0}\neq0$. An extension of this argument shows
that a generalized eigenvector with $j\geq2$ for $\lm$ cannot occur unless the
first $j-1$ derivatives of $\Lambda_1$ vanish at the root.

Bonilla {\em et al.} found that the transition between a bimodal distribution
$g(\om)$ supporting real eigenvalues and a bimodal distribution $g(\om)$
supporting complex eigenvalues occurs via a non-semisimple real eigenvalue
corresponding to a double root:
\be
\sfcn {1}{z_0}=0\hspace{0.75in}\sfcnp {1}{z_0}=0\hspace{0.75in}
\sfcnpp {1}{z_0}\neq0
\label{eq:dblrt}
\ee
where ${\rm Re}\,z_0=0$. For more complicated distributions, it is likely that
double roots with ${\rm Re}\,z_0\neq0$ are also possible. The case of a real
eigenvalue is considered here and we assume ${\rm Re}\,z_0=0$ for the remainder
of this section. The double root (\ref{eq:dblrt}) at $z_0$ then implies a
double root at $-z_0$ for $\Lambda_{-1}$:
\be
\sfcn {-1}{-z_0}=0\hspace{0.75in}\sfcnp {-1}{-z_0}=0\hspace{0.75in}
\sfcnpp {-1}{-z_0}\neq0.
\label{eq:dblrt2}
\ee

In applications with double roots, the frequency distribution will typically
depend on at least one additional parameter $\om_0$, for example in a bimodal
distribution $\om_0$ could be the distance between peaks in $g(\om,\omz)$. Then
$\Lambda_1'=0$ determines $z_0$ as a function of $K$ and $\omz$ and
$\Lambda_1=0$ implies a relation $K=K(\omz)$ between the parameters. The
eigenvector corresponding to $\sfcn {1}{z_0}=0$ is then $\ef=e^{i\th}\eff$
with
\be
\eff(\om)=\frac{-iK(\omz)/2}{(\om-z_0)-iD},\label{eq:efdr0}
\ee
and the generalized eigenvector $\gef=e^{i\th}\geff$ is determined by
(\ref{eq:gef5})
\be
\geff (\om)= \frac{K(\omz)/2}{(\om-z_0-iD)^2}.\label{eq:gef7}
\ee
One can always add to $v(\om)$ an arbitrary multiple of the homogeneous
solution $\eff$ but without loss of generality we have set the coefficient of
this term to zero. Reflection by $\k$ gives the eigenvector and generalized
eigenvector corresponding to the second root (\ref{eq:dblrt2}):
$\k\cdot\ef=e^{-i\th}\,\eff(-\om)$ and $\k\cdot\gef=e^{-i\th}\,\geff(-\om).$

For the adjoint eigenvalue problem, there is an eigenvector
$(\alop-\lm)\,\aef=0$ and generalized eigenvector $(\alop-\lm)\,\agef=\aef$
given by $\aef=e^{i\th}\,\aeff(\om)$ and $\agef=e^{i\th}\,\ageff(\om)$
where
\beq
\aeff(\om)&=& \left[\frac{-2\,g(\om,\omz)}{\sfcnpp{1}{z_0}(\om-z_0-iD)^2} +
\frac{2\,\sfcnppp{1}{z_0}g(\om,\omz)}
{3\sfcnpp{1}{z_0}^2(\om-z_0-iD)}\right]^\ast\\
\ageff(\om)&=&
\left[\frac{2i\,g(\om,\omz)}{\sfcnpp{1}{z_0}(\om-z_0-iD)}\right]^\ast.
\eeq
In addition the reflected vectors $\k\cdot\aef$ and $\k\cdot\agef$ solve the
same equations.
These eigenvectors satisfy the biorthogonality relations
\beq
(\aef,\ef)=<\aeff,\eff>=1&&
(\k\cdot\aef,\k\cdot\ef)=<\k\cdot\aeff,\k\cdot\eff>=1\\
(\agef,\gef)=<\ageff,\geff>=1&&
(\k\cdot\agef,\k\cdot\gef)=<\k\cdot\ageff,\k\cdot\geff>=1
\eeq
and
\beq
(\aef,\gef)=<\aeff,\geff>=0&&(\k\cdot\aef,\k\cdot\gef)=0\\
(\agef,\ef)=<\ageff,\eff>=0&&(\k\cdot\agef,\k\cdot\ef)=0.\\
\eeq

The evaluation of $P_\lambda\;\eta$ is now more involved because the poles are
not simple; from (\ref{eq:proj}) we have
\beq
P_\lambda\;\eta&=& \frac{e^{i\th}}{2\pi
i}\int_{\Gamma_0}\,d\lambda'\,R_1(\lm')\eta_1 +  \frac{e^{-i\th}}{2\pi
i}\int_{\Gamma_0}\,d\lambda'\,R_{-1}(\lm')\eta_{-1}\nonumber\\
&&\nonumber\\
&=& \frac{e^{i\th}}{2\pi i}\int_{\Gamma_0}\,d\lambda' \frac{K/2}
{\sfcn{1}{i\lm'}(i\om +D+\lm')}
\int^{\infty}_{-\infty}\,d\om'\,\frac{g(\om',\omz)\,\eta_1(\om')}{i\om'+D+\lm'}
\nonumber\\
&&+\frac{e^{-i\th}}{2\pi i}\int_{\Gamma_0}\,d\lambda' \frac{K/2}
{\sfcn{-1}{-i\lm'}(-i\om +D+\lm')} \int^{\infty}_{-\infty}\,d\om'\,
\frac{g(\om',\omz)\,\eta_{-1}(\om')}{-i\om'+D+\lm'}.
\label{eq:tbproj0}
\eeq
Consider the first integral for $K=K(\omz)$ when $\Lambda_1$ has a double root;
the contribution from this double pole is
\beq
\lefteqn{\frac{e^{i\th}}{2\pi i}\int_{\Gamma_0}\,d\lambda' \frac{K(\omz)/2}
{\sfcn{1}{i\lm'}(i\om +D+\lm')}
\int^{\infty}_{-\infty}\,d\om'\,\frac{g(\om',\omz)\,\eta_1(\om')}{i\om'+D+\lm'}}\\
&= &
e^{i\th}\frac{d}{d\lm'}
\left[ \frac{(\lm'-\lm)^2\,K(\omz)/2}{\sfcn{1}{i\lm'}(i\om+D+\lm')}
\left\{\int^{\infty}_{-\infty}\,d\om'\,\frac{g(\om',\omz)\,\eta_1(\om')}{i\om'
+D+\lm'}\right\} \right]_{\lm'=\lm}
\label{eq:dbl2}
\eeq
Expanding $\Lambda_1$,
\be
\sfcn{1}{i\lm'}=i^2\sfcnpp{1}{i\lm}\,\frac{(\lm'-\lm)^2}{2}+i^3
\sfcnppp{1}{i\lm}\,\frac{(\lm'-\lm)^3}{3!} +\cdots,\label{eq:dbl0}
\ee
and evaluating the right-hand side of (\ref{eq:dbl2}) gives
\be
\frac{e^{i\th}}{2\pi i}\int_{\Gamma_0}\,d\lambda' \frac{K(\omz)/2}
{\sfcn{1}{i\lm'}(i\om +D+\lm')}
\int^{\infty}_{-\infty}\,d\om'\,\frac{g(\om',\omz)\,\eta_1(\om')}{i\om'+D+\lm'}= (\aef,\eta)\ef+(\agef,\eta)\gef.\label{eq:dbl1}
\ee
A similar evaluation of the second integral in (\ref{eq:tbproj0}) yields
$(\aef,\eta)\ef+(\agef,\eta)\gef$ which is the complex conjugate of
(\ref{eq:dbl1}). Overall we have
\be
P_\lambda\;\eta=(\aef,\eta)\ef+(\agef,\eta)\gef+(\aef,\eta)\ef+(\agef,\eta)\gef;
\ee
thus a purely imaginary double root determines a real eigenvalue with two
eigenvectors and two generalized eigenvectors.

Complete expansions in terms of the eigenfunctions of $\flop l$ and hence
$\lop$ can be derived as in the Vlasov case\cite{ch} if desired.  We do not
need these expansions for the bifurcation analysis, it is sufficient to  know
how to project out the components of $\e$ along the critical eigenvectors in
the three cases of interest: semisimple real, semisimple complex, and
non-semisimple real.

\subsection{Example of a bimodal population}

\move We illustrate this discussion for the population
\be
g(\om,\omz)=\frac{\eps}{2\pi}\,\left[\frac{1}{(\om+\omz)^2+\eps^2} +
\frac{1}{(\om-\omz)^2+\eps^2}\right]\label{eq:bimode}
\ee
comprised of two sub-populations centered at $\om=\pm\omz$;
the $\eps\rightarrow0$ limit yields the delta function distribution considered
independently by Okuda and Kuramoto\cite{ok} and by Bonilla {\em et
al.}\cite{bns} When $\omz^2>\eps^2/3$, there are two peaks in $g(\om,\omz)$
located at
\be
\left(\frac{\om_{\rm peak}}{\omz}\right)^2=
\left[2-\sqrt{1+(\eps/\omz)^2}\,\right]\sqrt{1+(\eps/\omz)^2}.\label{eq:peaks}
\ee

For Im $z>-D$, the evaluation of $\Lambda_1$ yields\footnote{This case is
easily treated by closing the integration contour in the lower half plane. One
can show that for Im $z<-D$ and $K\geq0$, the spectral function has no roots.}
\be
\sfcn {1}{z}= \frac{[z+i(\eps+D)]^2-\omz^2-iK[z+i(\eps+D)]/2}
{[z+i(\eps+D)]^2-\omz^2}.\label{eq:bimodfcn}
\ee
which has two roots for $\omz>0$:
\be
z_\pm=i\,\left[(K/4)-\eps-D\pm\sqrt{(K/4)^2-\omz^2}\right].\label{eq:bimodrt}
\ee
At $\omz=0$, the expression in (\ref{eq:bimodfcn}) has only a single root
$z_+=i\,[(K/2)-\eps-D]$; the solution $z_-$ at $-i\,(\eps+D)$ is an artifact.
For $0<4\omz<K$, the roots are pure imaginary and the eigenvalues are real;
along the line $4\omz=K$ the roots collide, a double root forms, and a
non-semisimple real eigenvalue occurs. When $4\omz>K$, the eigenvalues are
complex and Hopf bifurcation is possible.

For $4\omz<K$, the root $z_+$ in (\ref{eq:bimodrt}) determines an eigenvalue if
Im $z_+>-D$ which requires either $K>2\,(\omz^2+\eps^2)/\eps$ or $K>4\eps$;
the solution $z_-$ determines an eigenvalue when
$4\eps<K<2\,(\omz^2+\eps^2)/\eps$. The eigenvalues $\lm_\pm=-i\,z_\pm$
disappear into the continuous spectrum at $-D$ when $\lm_\pm=-D$. This defines
a boundary
in parameter space
\be
K=K_{e}\equiv 2\,(\omz^2+\eps^2)/\eps
\ee
which is tangent to $K=4\omz$ at $\omz=\eps+D$.
The real eigenvalues cross the imaginary axis $\lm_\pm=0$ along a second
boundary
\be
K=K_{c}\equiv 2\,[\omz^2+(\eps^2+D^2)]/(\eps+D)\label{eq:kc}
\ee
with $\lm_+=0$ if $K<4(\eps+D)$ and $\lm_-=0$ if $K>4(\eps+D)$. This second
boundary is tangent to $K=4\omz$ at $K=4(\eps+D)$ where we find a
non-semisimple eigenvalue at zero; here the steady state bifurcation turns into
Hopf bifurcation. Figure 1 shows the $(K,\omz)$ parameter space with the curves
corresponding to eigenvalues coalescing with the continuum, crossing the
imaginary axis, or colliding on the real axis.

For $4\omz>K$, the roots (\ref{eq:bimodrt}) are complex and the eigenvalues
$(\lambda,\lambda^\ast)$ are
given by
\be
\lambda=-iz_0=\,(K-K_h)/4+ i\,\Omega \label{eq:hev}
\ee
where $\Omega\equiv\sqrt{\omz^2-(K/4)^2}$ and $K_h\equiv 4(\eps+D)$. If $K$
decreases below $K_e'\equiv 4\eps$ the eigenvalues $(\lambda,\lambda^\ast)$
disappear into the continuous spectrum at Re $\lm=-D$; as $K$ increases above
$K_h$ there is a Hopf bifurcation. The curves $K=K_e'$ and $K=K_h$ are also
shown in Fig. 1.

In an early discussion of bimodal populations without extrinsic noise $D=0$,
Kuramoto suggested that once the maximum of $g(\om)$ was no longer at the
average frequency (here $\overline{\om}=0$) the incoherent state would go
unstable for some $K=K_{c0}<K_c$ with synchronized behavior nucleating among
oscillators whose frequencies lay near the maxima of $g(\om)$.\cite{kur}
Although the development of bimodal peaks in (\ref{eq:bimode}) for
$\omz>\sqrt{\eps}/3$, does alter the stability properties of the incoherent
state; the effects are somewhat more complicated. As we shall see below in our
analysis of the steady state bifurcation, for
\be
(\eps+D)\,\sqrt{\frac{\eps+2D}{3\eps+4D}}<\omz<(\eps+D),
\ee
the incoherent state is stable when $K<K_c$ and there is a subcritical steady
state bifurcation to a synchronized state at $K=K_c$. There is no prior onset
of synchronization in the peaks of the distribution. The further qualitative
change in the stability properties of the incoherent state with increasing
$\omz$ above $(\eps+D)$ is to replace the steady state bifurcation by a Hopf
bifurcation which leads to time dependent behavior (cf. Section 3.2) in the
form of stable nonlinear ``standing waves''. These waves appear to be the type
of state Kuramoto envisioned; they are discussed in greater detail below. It
seems likely that these conclusions remain true in the limit $D\rightarrow0$.

\section{\hspace{0.125in}Codimension-one bifurcations}
\move For the nonlinear analysis to follow we require the Fourier expansion of
$\nop \e$. With
\be
e^{i\theta}\left(\e - i\frac{\partial\eta}{\partial \theta}\right)=
\sum_{l\neq 0,1}\,l\,\e_{l-1}(\om)\,e^{il\th}\label{eq:non3}
\ee
and
\be
\int^{2\pi}_0\,d\th'\int^{\infty}_{-\infty}\,d\om'
\eta(\th',\om',t)\,g(\om')\,e^{-i\th'}=2\pi\,<g,\e_1>,\label{eq:non4}
\ee
$\nop\e$ in (\ref{eq:nop}) becomes
\beq
\nop\e&=&-\pi K\left\{ e^{i\th}<g,\e_{-1}>\e_2 +
e^{-i\th}<g,\e_{1}>\e_{-2}\right.\nonumber\\
&&\hspace{0.4in}+\sum_{|l|\geq2}\,l\,e^{il\th} \left[<g,\e_{-1}>\,\e_{l+1}(\om)
- <g,\e_{1}>\,\e_{l-1}(\om)\right]\}.\label{eq:non5}
\eeq

\subsection{Steady-state bifurcation}

\move We consider a simple pure imaginary root $\sfcn{1}{z_0}=0$ so that
$\lm=-iz_0$ is real. As a parameter such as $K$ is varied,  we assume that
$\lm$ crosses through $\lm=0$ at $K=K_c$. From (\ref{eq:efa}) -
(\ref{eq:efacc}) there are two eigenfunctions $\ef$ and $\k\cdot\ef=\ef^\ast$,
with amplitudes $(\alpha,\alpha^\ast)$ defined by
$\alpha(t)\equiv(\aef,\eta).$ The distribution $\eta$ decomposes into the
critical linear modes
$P_\lambda\;\eta=\alpha(t)\ef(\th,\om)+cc,$
and the remaining degrees of freedom
$S(\th,\om)\equiv\eta-P_\lambda\;\eta.$
Thus we have
\be
\e(\th,\om,t)=[\alpha(t)\ef(\th,\om)+ cc] + S(\th,\om,t)\label{eq:mode1}
\ee
where $(\aef,S)=0.$
In terms of the Fourier components of $\e$ and $S$, we have
\be
\e_l=\alpha\,\eff\,\delta_{l,1}+\alpha^\ast\,\eff^\ast\,\delta_{l,-1} +
S_l;\label{eq:el}
\ee
note that the normalization (\ref{eq:mean}) requires
\be
S_{l=0}=0.\label{eq:s0}
\ee

By projecting with $\aef$, the evolution equation (\ref{eq:dyn}) can be
rewritten as
\beq
\dot{\alpha}&=&\lm\, \alpha +(\aef,\nop{\eta}) \label{eq:alph1}\\
\frac{\partial S}{\partial t}&=&\lop S+\left\{\nop\e
-\left[(\aef,\nop{\eta})\,\ef  + cc \right]\right\};\label{eq:S1}
\eeq
since $<g,\e_l>= \alpha\,\delta_{l,1}+\alpha^\ast\,\delta_{l,-1} + <g,S_l>$
in (\ref{eq:non5}), the Fourier decomposition of these equations gives
\beq
\dot{\alpha}&=&\lm \alpha-\pi K\,(\alpha^\ast +
<g,S_{-1}>)<\aeff,S_2>\label{eq:alph2}\\
\frac{\partial S}{\partial t}&=&\lop S -\pi K
\left\{ e^{i\th}(\alpha^\ast+<g,S_{-1}>)(S_2-<\aeff,S_2>\eff)\right.\nonumber\\
&&\hspace{0.6in} +2\,e^{i2\th}\left[(\alpha^\ast + <g,S_{-1}>)S_{3}(\om) -
(\alpha+ <g,S_{1}>)(\alpha\,\eff+S_1)\right]\nonumber\\
&&\hspace{0.6in}+\sum_{l\geq3}\,l\,e^{il\th}
\left[(\alpha^\ast + <g,S_{-1}>)S_{l+1}(\om) -
(\alpha+ <g,S_{1}>)S_{l-1}(\om)\right]\nonumber\\
&&\left.\hspace{2.6in}+cc\right\}.\label{eq:S2}
\eeq

\subsubsection{Dynamics on the center manifold}

\move The equations (\ref{eq:alph2})-(\ref{eq:S2}) are obviously coupled but on
the center manifold the ``transverse'' degrees of freedom $S$ depend on time
through the critical amplitudes; this may be expressed in terms of a function
$H(\th,\om,\alpha,\alpha^\ast)$:
\be
S(\th,\om,t)=H(\th,\om,\alpha(t),\alpha(t)^\ast)=
\sum_{l=-\infty}^{\infty}H_l(\th,\alpha(t),\alpha(t)^\ast)\,e^{il\th}.
\label{eq:scm}
\ee
In geometric terms, $H$ describes the graph of the center manifold near the
incoherent state.\cite{craw5,gh,iv} Thus solutions on the center manifold
satisfy (\ref{eq:alph2}) - (\ref{eq:S2}) with  $S$ replaced by $H$
\beq
\dot{\alpha}&=&\lm \alpha-\pi K\,(\alpha^\ast + <g,H_{-1}>)<\aeff,H_2>
\label{eq:2dcm0}\\
\frac{\partial S}{\partial t}&=&\lop H -\pi K
\left\{ e^{i\th}(\alpha^\ast+<g,H_{-1}>)(H_2-<\aeff,H_2>\eff)\right.\nonumber\\
&&\hspace{0.6in} +2\,e^{i2\th}\left[(\alpha^\ast + <g,H_{-1}>)H_{3}(\om) -
(\alpha+ <g,H_{1}>)(\alpha\,\eff+H_1)\right]\nonumber\\
&&\hspace{0.6in}+\sum_{l\geq3}\,l\,e^{il\th}
\left[(\alpha^\ast + <g,H_{-1}>)H_{l+1}(\om) -
(\alpha+ <g,H_{1}>)H_{l-1}(\om)\right]\nonumber\\
&&\left.\hspace{2.6in}+cc\right\};\label{eq:S2cm}
\eeq
the equation (\ref{eq:2dcm0}) for the mode amplitude now decouples from $S$ and
defines an autonomous two-dimensional system.

The center manifold dynamics (\ref{eq:2dcm0}) and the graph function are both
constrained by the $\otwo$ symmetry of the problem.  The Fourier coefficients
of $H$ for $l>0$ must have the form
\be
H_l(\om,\alpha,\alpha^\ast)=\alpha^l\,h_l(\om,\sigma);\label{eq:rothl}
\ee
where $h_l(\om,\sigma)$ satisfies
\be
h_l(-\om,\sigma)^\ast=h_l(\om,\sigma)\label{eq:refhl}
\ee
and $\sigma=|\alpha|^2$ denotes the basic $\otwo$ invariant. Similarly
(\ref{eq:2dcm0}), the equation for $\alpha$ must have the general form
\be
\dot{\alpha}=p(\sigma)\alpha\label{eq:cmnf1}
\ee
where $p(\sigma)$ is a real-valued function of a single variable. For small
$\alpha$, the leading terms in (\ref{eq:2dcm0}) are therefore
\be
\dot{\alpha}=\alpha[p(0)+p_{\sigma}(0)|\alpha|^2+\cdots]\label{eq:cmnf2}
\ee
and comparison to (\ref{eq:2dcm0}) shows $p(0)=\lm$ and $p_{\sigma}(0)$ remains
to be calculated.

We briefly summarize the derivation of (\ref{eq:rothl}) - (\ref{eq:cmnf1}).
The form of $H$ is constrained by the fact that the center manifold is mapped
to itself $\otwo$ transformations.\cite{ruelle} More precisely, if
$\e(\th,\om)$ corresponds to a point on the center manifold and  $\g\in\otwo$,
then the transformed function $\g\cdot\e$ will also lie on the center manifold.
When we represent elements of the center manifold using (\ref{eq:scm}),
\be
\e(\th,\om)=\alpha\,\ef(\th,\om)+\alpha^\ast\,\ef^\ast(\th,\om)+H(\th,\om,\alpha,\alpha^\ast),\label{eq:cmfcn}
\ee
then $\g\cdot\e$ must also be of this form
\be
(\g\cdot\e)(\th,\om)=\alpha'\,\ef(\th,\om)+{\alpha'}^\ast\,\ef^\ast(\th,\om)+H(\th,\om,\alpha',{\alpha'}^\ast)\label{eq:cmfcno2}
\ee
for some transformed amplitude $\alpha'$.
The constraints on $H$  implied by this requirement follow from considering the
generators of $\otwo$ in (\ref{eq:phi}) - (\ref{eq:kap}). In conjunction with
(\ref{eq:rep}) we find for rotations
\be
[\alpha\,\ef(\th-\phi,\om)+cc]+H(\th-\phi,\om,\alpha,\alpha^\ast)=
[\alpha\,e^{-i\phi}\,\ef(\th,\om)+cc]+
H(\th,\om,e^{-i\phi}\,\alpha,e^{i\phi}\,\alpha^\ast)
\ee
and
\be
[\alpha\,\ef(-\th,-\om)+cc]+H(-\th,-\om,\alpha,\alpha^\ast)=
[\alpha^\ast\,\ef(\th,\om)+cc]+ H(\th,\om,\alpha^\ast,\alpha)
\ee
for reflection.  Since $\ef(\th-\phi,\om)=e^{-i\phi}\,\ef(\th,\om)$ and
$\ef(-\th,-\om)=\ef(\th,\om)^\ast$, the amplitude $\alpha$ transforms under
$\otwo$ by
\beq
\phi\cdot\alpha&=&e^{-i\phi}\alpha\label{eq:rotcm}\\
\k\cdot\alpha&=&\alpha^\ast,\label{eq:refcm}
\eeq
and $H$ must satisfy
\beq
H(\th-\phi,\om,\alpha,\alpha^\ast)&=&
H(\th,\om,e^{-i\phi}\,\alpha,e^{i\phi}\,\alpha^\ast)\label{eq:rotH}\\
H(-\th,-\om,\alpha,\alpha^\ast)&=&
H(\th,\om,\alpha^\ast,\alpha).\label{eq:refH}
\eeq
Restated for the Fourier coefficients $H_l$, these relations require
\be
e^{-il\phi}H_l(\om,\alpha,\alpha^\ast)=
H_l(\th,\om,e^{-i\phi}\,\alpha,e^{i\phi}\,\alpha^\ast)\label{eq:rotHl}
\ee
and
\be
H_l(-\om,\alpha,\alpha^\ast)^\ast= H_l(\om,\alpha^\ast,\alpha),\label{eq:refHl}
\ee
respectively. The rotational constraint (\ref{eq:rotHl}) states that the
function $(\alpha^\ast)^l\,H_l$  is invariant under rotations, hence it depends
on the mode amplitudes $(\alpha,\alpha^\ast)$  only through the basic
invariant\footnote{cf. Golubitsky, Stewart, and Schaeffer, chapter XII, \S 4}
$\sigma=|\alpha|^2$ and must contain an overall factor of $(\alpha^\ast)^l$:
\be
(\alpha^\ast)^l\,H_l(\om,\alpha,\alpha^\ast)=\sigma^l\,h_l(\om,\sigma)\label{eq:rotinv}
\ee
where $h_l(\om,\sigma)$ is an arbitrary function of $\om$ and $\sigma$.
Dividing by $(\alpha^\ast)^l$ gives the form of $H_l$ in (\ref{eq:rothl});
the reflection symmetry (\ref{eq:refHl}) then further requires that
$h_l(\om,\sigma)$ satisfy (\ref{eq:refhl}).

The $\otwo$ invariance of the center manifold also implies that the vector
field in (\ref{eq:2dcm0}) must commute with the representation of $\otwo$ given
in (\ref{eq:rotcm}) - (\ref{eq:refcm}). This can be explicitly verified using
the properties of $H_l$ in (\ref{eq:rotHl}) - (\ref{eq:refHl}). Any smooth
two-dimensional vector field that commutes with this representation must have
the form (\ref{eq:cmnf1}); see for example Golubitsky, Stewart, and Schaeffer,
chapter XII, \S 5.

\subsubsection{Evaluation of $p_{\sigma}(0)$}

\move From (\ref{eq:rothl}), $H_{1}$ and $H_2$ have the form
\beq
H_{1}&=&\alpha\,h_1(\om,\sigma)=\alpha\,[h_1^{(0)}(\om)+\ord
{\sigma}]\label{eq:Hl1}\\
H_2 &=&\alpha^2\,h_2(\om,\sigma)=\alpha^2\,[h_2^{(2)}(\om)+
\ord{\sigma}],\label{eq:Hl2}
\eeq
however the geometry of the center manifold requires that the lowest order
terms be quadratic in $(\alpha,\alpha^\ast)$ so $h_{1}^{(0)}=0$; hence the
center manifold vector field (\ref{eq:2dcm0}) becomes
\beq
\dot{\alpha}&=&\lm \alpha-\pi K\,\alpha\,|\alpha|^2\,(1
+<g,h_{1}^\ast>)<\aeff,h_2>\label{eq:2dcm}\\
&=&\alpha[\lm -\pi K\,<\aeff,h_2^{(2)}>|\alpha|^2 +\cdots]. \label{eq:2dcmnf}
\eeq
Comparing (\ref{eq:cmnf2}) and (\ref{eq:2dcmnf}) we obtain $p_{\sigma}(0)$
\be
p_{\sigma}(0)=-\pi K\, <\aeff,h_2^{(2)}>\label{eq:cub1}
\ee
in terms of the center manifold coefficient $h_2^{(2)}$.

When $S$ is given by (\ref{eq:scm})
we have
\be
\frac{\partial S}{\partial t}=\frac{\partial H}{\partial \alpha}\dot{\alpha} +
\frac{\partial H}{\partial \alpha^\ast}\dot{\alpha}^\ast
\ee
for solutions on the center manifold,
and consistency with (\ref{eq:S2}) requires
\beq
\frac{\partial H}{\partial \alpha}\dot{\alpha} + \frac{\partial H}{\partial
\alpha^\ast}\dot{\alpha}^\ast&=&
\lop H -\pi K
\left\{ e^{i\th}
(\alpha^\ast+<g,H_{-1}>)(H_2-<\aeff,H_2>\eff)\right.\label{eq:Heqn}\\
&&\hspace{0.6in}
+2\,e^{i2\th} \left[(\alpha^\ast + <g,H_{-1}>)H_{3}(\om) -
(\alpha+ <g,H_{1}>)(\alpha\,\eff+H_1)\right]\nonumber\\
&&\hspace{0.4in}+\sum_{l\geq3}\,l\,e^{il\th} \left[
(\alpha^\ast + <g,H_{-1}>)H_{l+1}(\om) -
(\alpha+ <g,H_{1}>)H_{l-1}(\om)\right]\nonumber\\
&&\left.\hspace{2.6in}+cc\right\}\nonumber
\eeq
where $\dot{\alpha}$ is given by (\ref{eq:2dcm0}). This provides the equation
needed to determine the expansion coefficients of $H$;
for small $|\alpha|$, we can solve (\ref{eq:Heqn}) using a power series
expansion in $(\alpha,\alpha^\ast)$. Separating the different Fourier
components of (\ref{eq:Heqn}) for $l=1$, $l=2$ and $l\geq3$, we find
\beq
\frac{\partial H_1}{\partial \alpha}\dot{\alpha}+\frac{\partial H_1}{\partial
\alpha^\ast}\dot{\alpha}^\ast&=&
\flop 1H_1 -\pi K (\alpha^\ast+<g,H_{-1}>)(H_2-<\aeff,H_2>\eff)
\label{eq:H1}\\
&&\nonumber\\
\frac{\partial H_2}{\partial \alpha}\dot{\alpha}+\frac{\partial H_2}{\partial
\alpha^\ast}\dot{\alpha}^\ast&=&
\flop 2H_2 -2\pi K \,\left[(\alpha^\ast + <g,H_{-1}>)H_{3}(\om)
\right.\nonumber\\
&&\left.\hspace{1.5in}
- (\alpha+ <g,H_{1}>)(\alpha\,\eff+H_1)\right]\label{eq:H2}\\
&&\nonumber\\
\frac{\partial H_l}{\partial \alpha}\dot{\alpha}+\frac{\partial H_l}{\partial
\alpha^\ast}\dot{\alpha}^\ast&=&
\flop lH_l  -l\pi K \left[(\alpha^\ast +
<g,H_{-1}>)H_{l+1}(\om)\right.\nonumber\\
&&\left.\hspace{1.5in}- (\alpha+ <g,H_{1}>)H_{l-1}(\om)\right]
\label{eq:Hl}
\eeq
respectively, and inspecting these equations shows that the inhomogeneous term
$\alpha^2\eff$ in (\ref{eq:H2}) that forces a non-zero term $h_2^{(2)}(\om)$ at
the lowest order term in $H_2$. Since $h_1^{(0)}(\om)=0$, the balance at second
order in (\ref{eq:H2}) is $2\lm\,h_2^{(2)}=\flop 2 h_2^{(2)} + 2\pi\,K\,\eff$
which gives
\be
h_2^{(2)}(\om)=\frac{-i\pi K\,\eff(\om)}{\om-z_0-i2D}.\label{eq:cmcoeff}
\ee

Thus the cubic coefficient in the center manifold dynamics (\ref{eq:cub1}) is
given by
\be
p_{\sigma}(0)=\frac{-\pi^2 \,K^3}{2\,\sfcnp{1}{z_0}}
\int^{\infty}_{-\infty}\,d\om\,\frac{g(\om)} {(\om-z_0-iD)^2\,(\om-z_0-i2D)}.
\label{eq:nfcoeff}
\ee
Note that $p_{\sigma}(0)$ is real as expected from $\otwo$ symmetry.

\subsubsection{Example of a bimodal population}

For the population in (\ref{eq:bimode}) the first steady state bifurcation
occurs when the eigenvalue $\lm_+$crosses zero; this occurs along $K=K_c$ Fig.
1. Along this steady state locus, the coefficient (\ref{eq:nfcoeff}) can be
easily evaluated by closing the contour in the lower half plane:
\be
p_{\sigma}(0)=\frac{\pi^2\,K_c^2\,(3\eps+4D)\,(\omz^2-\om_d^2)}
{2\,[(\eps+D)^2-\omz^2]\,[(\eps+2D)^2+\omz^2]}
\ee
where
\be
\om_d\equiv(\eps+D)\,\sqrt{\frac{\eps+2D}{3\eps+4D}}.
\ee
Since $\omz<(\eps+D)$ when this bifurcation occurs, the overall sign of
$p_{\sigma}(0)$ is determined by $\omz^2-\om_d^2$. For $\omz<\om_d$, the
bifurcation is supercritical to a stable synchronized state: for $\omz>\om_d$,
we have $p_{\sigma}(0)>0$ and the bifurcation is subcritical. Note that
$p_{\sigma}(0)$ is finite in the limit $D\rightarrow0$ as long as $\eps>0$;
this feature is examined in greater generality below.

For $\eps>0$ and $D=0$, the transition from supercritical to subcritical
bifurcation was noted by Kuramoto,\cite{kur} and  at $\eps=0$ and $D>0$ it was
found by Bonilla {\em et al.} and Okuda and Kuramoto.\cite{bns,ok} The former
group also calculated the fifth order term necessary to determine that the
subcritical branch turns around at a saddle node bifurcation and becomes stable
at finite amplitude.

\subsubsection{The $D\rightarrow0$ limit}

\move If $D=0$ then the critical eigenvalues emerge from the continuous
spectrum simultaneously with the onset of the linear instability, and it is of
interest to examine the behaviour of the amplitude equation in this regime. At
criticality for either steady-state or Hopf bifurcation Re$\lm=0$ so Im$z_0=0$,
thus we consider the behavior of $p_{\sigma}(0)$ when $D\rightarrow0$ with
$z_0=0$. The integrand in (\ref{eq:nfcoeff}) has  poles at $\om=iD$ and
$\om=i2D$ that approach the contour of integration as $D\rightarrow0$. We
assume $g(\om)$ is analytic at $\om=0$ so that the contour can be deformed
slightly below the axis and the limits evaluated using the Plemej formulas.

It is convenient to first consider the limiting behavior of $\Lambda_l$. From
(\ref{eq:Lam}) we re-express the derivatives of $\sfcn l{z}$ by integrating by
parts
\be
\frac{\partial^n}{\partial z^n}\sfcn l{z}
=\delta_{n,0}+ \frac{ilK}{2} \int^{\infty}_{-\infty}\,d\om\frac{\partial^n
g(\om)/\partial\om^n}{(\om -z-ilD)};\label{eq:lamd1}
\ee
thus when $z=r$ is real-valued, then as $D\rightarrow0$
\be
\lim_{D\rightarrow0}\frac{\partial^n}{\partial z^n}\sfcn 1{r}=
\delta_{n,0}+\frac{iK}{2}\left[{\cal P}
\int^{\infty}_{-\infty}\,d\om\frac{\partial^n g(\om)/\partial\om^n} {\om-r} +
i\pi\,\frac{\partial^n g(r)}{\partial\om^n}\right]\label{eq:lmd1}
\ee
where ${\cal P}$ denotes a principal value integral.
For steady-state bifurcation, when $g(\om)$ is even in $\om$ and $r=0$ at
$K=K_c$, these limits simplify:
\beq
\lim_{D\rightarrow0}\frac{\partial^n}{\partial z^n}\sfcn 1{0}&=&
{iK_c}\int^{\infty}_{0}\,d\om\frac{\partial^n g(\om)/\partial\om^n}{\om}
\hspace{0.3in}n\;\;\mbox{\rm odd}\label{eq:lmd3}\\
&&\nonumber\\
\lim_{D\rightarrow0}\frac{\partial^n}{\partial z^n}\sfcn 1{0}&=&
\delta_{n,0}+\frac{-\pi\,K_c}{2}\,\frac{\partial^n g(0)}{\partial\om^n}
\hspace{0.3in}n\;\;\mbox{\rm even}.\label{eq:lmd4}
\eeq
Note that for $n=0$ by setting $\sfcn 1{0}=0$ in (\ref{eq:lmd4}) recovers the
previous result (\ref{eq:kcrit}) for $K_c=2/\pi g(0)$.

For small $D$ the behavior of $p_{\sigma}(0)$ is extracted in the same way;
setting $z_0=0$ we find
\beq
p_{\sigma}(0)&=&\frac{-\pi^2\, K_c^3}{2\,\sfcnp{1}{0}}\,\lim_{D\rightarrow0}
\int^{\infty}_{-\infty}\,d\om\,\frac{g(\om)} {(\om-iD)^2\,(\om-i2D)}\nonumber\\
&=&i\,(\pi K_c)^2\frac{\sfcnpp{1}{0}}{2\sfcnp{1}{0}} +\ord D\nonumber\\
&=&\frac{-\pi }{g(0)^2}\,\frac{\partial^2 g(0)}{\partial\om^2}\,
\left[\int^{\infty}_{0}\,d\om\frac{\partial
g(\om)/\partial\om}{\om}\right]^{-1} +\ord D.\label{eq:limit}
\eeq
Once differences in notation are taken into account, this result agrees with
the coefficient obtained by Kuramoto by the ``self-consistent''
method.\cite{kur}
Notably absent is the kind of singular behavior found in the Vlasov amplitude
equations in the corresponding limit.\cite{jdc} For a monotonic
reflection-symmetric profile this gives $p_{\sigma}(0)<0$ in the limit of weak
diffusion. Thus at the level of the amplitude equation, we expect supercritical
bifurcation to a stable synchronized state in this regime. However when $D=0$
center manifold theory no longer justifies our reduction to two dimensions; the
qualitative agreement at $D=0$ between numerical simulations\cite{kur84} and
our amplitude equation may be fortuitous.

\subsubsection{Perturbing $\otwo\rightarrow\sotwo$}

\move If the distribution $g(\om)$ is perturbed so that the reflection symmetry
(\ref{eq:g}) is broken, then the bifurcation problem (in the rotating frame
(\ref{eq:rotf})) has only the rotational symmetry $\sotwo$. In this case one
expects the real eigenvalue of multiplicity-two to split into a non-degenerate
complex conjugate pair and the steady state bifurcation considered here to be
perturbed to a Hopf bifurcation.  The resulting Hopf bifurcation leads to
time-periodic states in the form of rotating or travelling waves; Knobloch
discusses such perturbed bifurcations in more detail.\cite{kno}

\subsection{Hopf bifurcation}

\move We next consider a simple complex root $\sfcn{1}{z_0}=0$ so that
$z_0\neq-z_0^\ast$ and $\lm=-iz_0$ is complex. As $K$ is varied through
$K_h$,we assume that ${\rm Re}\,\lm$ crosses through ${\rm Re}\,\lm=0$ at
$K=K_h$.
{}From (\ref{eq:hopfev2}) there are two eigenfunctions $\ef$ and
$\k\cdot\ef=\ef^\ast$, with amplitudes $(\alpha,\beta)$ defined by
$\alpha(t)\equiv(\aef,\eta)$ and
$\beta(t)\equiv(\k\cdot\aef,\eta).$
With the projections (\ref{eq:hev1}) - (\ref{eq:hev2}), the distribution $\eta$
decomposes as before,
\be
\e(\th,\om,t)=[\alpha(t)\;\ef(\th,\om)+\beta(t)\;\k\cdot\ef(\th,\om)+ cc] +
S(\th,\om,t)\label{eq:modehopf}
\ee
where $(\aef,S)=(\k\cdot\aef,S)=0.$

Projecting (\ref{eq:dyn}) with $\aef$ and $\k\cdot\aef$ yields
\beq
\dot{\alpha}&=&\lm \alpha +(\aef, \nop{\eta})\label{eq:a1}\\
\dot{\beta}&=&\lm \beta +(\k\cdot\aef, \nop{\eta})\label{eq:b1}\\
\frac{\partial S}{\partial t}&=&\lop S+\left\{\nop\e -
\left[\ef\;(\aef, \nop{\eta})+\k\cdot\ef\; (\k\cdot\aef, \nop{\eta})+cc\right]
\right\}.\label{eq:st1}
\eeq
In terms of the Fourier components of $\e_l$
\beq
\e_l&=&(\alpha\,\eff+ \beta^\ast\,\k\cdot\eff^\ast)\,\delta_{l,1}+
(\alpha^\ast\,\eff^\ast+ \beta\,\k\cdot\eff)\,\delta_{l,-1}+S_l
\label{eq:fcoeff1}\\
<g,\e_l>&=& (\alpha+ \beta^\ast)\,\delta_{l,1}+ (\alpha^\ast+
\beta)\,\delta_{l,-1} + <g,S_l>,\label{eq:fcoeff2}
\eeq
and the expansion of $\nop\e$ in (\ref{eq:non5}), (\ref{eq:a1})-(\ref{eq:st1})
can be re-written as
\beq
\dot{\alpha}&=&\lm \alpha-\pi K\,(\alpha^\ast +\beta
+<g,S_{-1}>)<\aeff,S_2>\label{eq:a2}\\
\dot{\beta}&=&\lm \beta-\pi K\,(\alpha +\beta^\ast
+<g,S_{1}>)<\k\cdot\aeff,S_{-2}>\label{eq:b2}\\
\frac{\partial S}{\partial t}&=&\lop S -\pi K
\left\{ e^{i\th} (\alpha^\ast +\beta +<g,S_{-1}>)
[S_2-<\aeff,S_2>\eff-<\k\cdot\aeff^\ast,S_2>\k\cdot\eff^\ast]\right.\nonumber\\
&&\hspace{0.6in}
+2\,e^{i2\th} \left[(\alpha^\ast +\beta+ <g,S_{-1}>)S_{3}(\om) -
(\alpha+\beta^\ast+ <g,S_{1}>)(\alpha\,\eff+ \beta^\ast\,\k\cdot\eff^\ast
+S_1)\right]\nonumber\\
&&\hspace{0.4in}+\sum_{l\geq3}\,l\,e^{il\th} \left[
(\alpha^\ast + \beta+ <g,S_{-1}>)S_{l+1}(\om) -
(\alpha+ \beta^\ast+ <g,S_{1}>)S_{l-1}(\om)\right]\nonumber\\
&&\hspace{2.6in} + cc\}.\label{eq:st2}
\eeq

\subsubsection{Dynamics on the center manifold}

\move For a solution on the center manifold near the incoherent state, we can
express the time-dependence of $S(\th,\om,t)$ as
\be
S(\th,\om,t)=H(\th,\om,\alpha(t),\alpha(t)^\ast,\beta(t),\beta(t)^\ast)
=\sum_{l=-\infty}^{\infty}
\,H_l(\om,\alpha(t),\alpha(t)^\ast,\beta(t),\beta(t)^\ast)\, e^{il\th}
\label{eq:scmh}
\ee
and restrict (\ref{eq:a2}) - (\ref{eq:st2}) to these solutions by substituting
$S=H$. This determines the autonomous four-dimensional system:
\beq
\dot{\alpha}&=&\lm \alpha-\pi K\,(\alpha^\ast +\beta
+<g,H_{-1}>)<\aeff,H_2>\label{eq:ahopf}\\
\dot{\beta}&=&\lm \beta-\pi K\,(\alpha +\beta^\ast
+<g,H_{1}>)<\k\cdot\aeff,H_{-2}>.\label{eq:bhopf}
\eeq

For this bifurcation the mode amplitudes transform according to
\beq
\phi\cdot(\alpha,\beta)&=&(e^{-i\phi}\alpha,e^{i\phi}\beta)\label{eq:rotcmh}\\
\k\cdot(\alpha,\beta)&=&(\beta,\alpha),\label{eq:refcmh}
\eeq
and the $\otwo$-invariance of the center manifold implies that the Fourier
coefficients of $H$ have the form
\be
H_l= \sum_{j=0}^l\,\alpha^{j}\,(\beta^\ast)^{l-j}\,
H_l^{(j)}(\om,|\alpha|^2,|\beta|^2,\alpha\beta,\alpha^\ast\beta^\ast)
\label{eq:hopfhfc}
\ee
for $l>0$. Here $H_l^{(j)}$ are functions of
$|\alpha|^2,|\beta|^2,\,\alpha\beta,\alpha^\ast\beta^\ast$ and $\om$ that
satisfy
\be
H_l^{(j)}(-\om,|\alpha|^2,|\beta|^2,\alpha\beta,\alpha^\ast\beta^\ast)^\ast =
H_l^{(l-j)}(\om,|\beta|^2,|\alpha|^2,\alpha\beta,\alpha^\ast\beta^\ast)
\label{eq:hrefHlc}
\ee
for $j=1,2,\ldots,l$. The analysis leading to (\ref{eq:hopfhfc}) -
(\ref{eq:hrefHlc}) is analogous to the argument given above for steady-state
bifurcation: the invariance of the manifold requires the Fourier components of
$H$ to satisfy
\be
e^{-il\phi}H_l(\om,\alpha,\alpha^\ast,\beta,\beta^\ast)=
H_l(\th,\om,e^{-i\phi}\,\alpha,e^{i\phi}\,\alpha^\ast,
e^{i\phi}\,\beta,e^{-i\phi}\,\beta^\ast), \label{eq:hrotHl}
\ee
and
\be
H_l(-\om,\alpha,\alpha^\ast,\beta,\beta^\ast)^\ast=
H_l(\om,\beta,\beta^\ast,\alpha,\alpha^\ast),\label{eq:hrefHl}
\ee
The constraint (\ref{eq:hrotHl}) implies that the function
$(\alpha^\ast)^l\,H_l$  is invariant under rotations, hence it can only depend
on the mode amplitudes only through the basic rotation invariants
$|\alpha|^2,|\beta|^2,\alpha\beta,$ and $\alpha^\ast\beta^\ast$ and must
contain an overall factor of $(\alpha^\ast)^l$. By combining $|\alpha|^2$ and
$\alpha^\ast\beta^\ast$ there are $l+1$ distinct ways to generate this factor
so $(\alpha^\ast)^l\,H_l$ takes the form
\be
(\alpha^\ast)^lH_l=
\sum_{j=0}^l\,|\alpha|^{2j}\,(\alpha^\ast\beta^\ast)^{l-j}\,
H_l^{(j)}(\om,|\alpha|^2,|\beta|^2,\alpha\beta,\alpha^\ast\beta^\ast)
\ee
where the functions $H_l^{(j)}$ are arbitrary functions of
$|\alpha|^2,|\beta|^2,\,\alpha\beta,\alpha^\ast\beta^\ast$ and $\om$. Thus
$H_l$ has the asserted form (\ref{eq:hopfhfc}); the condition imposed by
reflection invariance (\ref{eq:hrefHl}) implies (\ref{eq:hrefHlc}).

For $l=1$ and $l=2$, the components needed in (\ref{eq:ahopf}) and
(\ref{eq:bhopf}), we have
\beq
H_1&=&
\alpha\,H_1^{(1)}(\om,|\alpha|^2,|\beta|^2,\alpha\beta,\alpha^\ast\beta^\ast) +
\beta^\ast\,
H_1^{(0)}(\om,|\alpha|^2,|\beta|^2,\alpha\beta,\alpha^\ast\beta^\ast)\\
H_2&=&
\alpha^2\,H_2^{(2)}(\om,|\alpha|^2,|\beta|^2,\alpha\beta,\alpha^\ast\beta^\ast)
+ \alpha\,\beta^\ast\,
H_2^{(1)}(\om,|\alpha|^2,|\beta|^2,\alpha\beta,\alpha^\ast\beta^\ast) +
\nonumber\\
&&\hspace{1.0in} (\beta^\ast)^2\,
H_2^{(0)}(\om,|\alpha|^2,|\beta|^2,\alpha\beta,\alpha^\ast\beta^\ast).
\eeq
Since the center manifold geometry requires that the Taylor expansion of $H$
begins at second order, the leading order terms in these components are
\beq
H_1&=& \ord 3\\
H_2&=&\alpha^2\,h_2^{(2)}(\om)
+\alpha\beta^\ast\,h_2^{(1)}(\om)+(\beta^\ast)^2\,h_2^{(0)}(\om) +\ord 3;
\label{eq:h2hopf}
\eeq
the third order terms will not be required to determine the cubic terms in the
amplitude equations. Note that the general relation (\ref{eq:hrefHlc}) implies
\beq
h_2^{(2)}(-\om)^\ast&=&h_2^{(0)}(\om)\label{eq:h0id}\\
h_2^{(1)}(-\om)^\ast&=&h_2^{(1)}(\om).\label{eq:h1id}
\eeq

The four-dimensional vector field (\ref{eq:ahopf})-(\ref{eq:bhopf}) commutes
with the $\otwo$ symmetry (\ref{eq:rotcmh}) - (\ref{eq:refcmh}) on the center
manifold; in addition this system can be further simplified by making a
nonlinear change of variables $(\alpha,\beta)\rightarrow(\alpha',\beta')$ to
put (\ref{eq:ahopf})-(\ref{eq:bhopf}) in {\pbnf}.\cite{gss,craw} This change of
variables introduces an additional ${\rm S}^1$ symmetry defined by
$(\alpha,\beta)\rightarrow e^{i\chi}(\alpha,\beta);$
this is the ``phase shift symmetry'' that characterizes the Hopf normal form
even in bifurcations without symmetry.\cite{craw5,gh}
Thus the normal form for (\ref{eq:bhopf}) will have $\otwoh$ symmetry and must
be of the general form\cite{gss,craw}
\be
\left(\begin{array}{c}\dot{\alpha'}\\\dot{\beta'}\end{array}\right) =
\bigl[p(u,\Delta)+iq(u,\Delta)\bigr]
\left(\begin{array}{c} {\alpha'}\\{\beta'}\end{array}\right)
+\bigl[r(u,\Delta)+is(u,\Delta)\bigr]\delta
\left(\begin{array}{c} {\alpha'}\\{-\beta'}\end{array}\right),
\label{eq:hopfnf0}
\ee
where $\delta=|\beta'|^2-|\alpha'|^2$ and $p,q,r,$ and $s$ are
real-valued functions of the two basic invariants $u=|\alpha'|^2+
|\beta'|^2$ and $\Delta=\delta^2$.
A final reduction to two dimensions is accomplished by introducing amplitudes
and phases $\alpha'=x_1 e^{i\sigma_1}$ and $\beta'=x_2 e^{i\sigma_2}$ into
(\ref{eq:hopfnf0}) and
noting that the two-dimensional amplitude equations,
\beq
\dot x_1 &= [p+\delta r]x_1\label{eq:d4nf0}\\
\dot x_2 &= [p-\delta r]x_2\label{eq:d4nf1}
\eeq
are decoupled from the phase evolution $\dot\sigma_1= q+\delta s$ and
$\dot\sigma_2= q-\delta s.$
This decoupling is a consequence of the phase shift symmetry. Upon expanding
(\ref{eq:d4nf0}) and (\ref{eq:d4nf1}) we obtain the leading order behavior
\beq
\dot x_1 &=[p(0)+(p_u(0)-r(0))x_1^2+(p_u(0)+r(0))x_2^2+\cdots]x_1
\label{eq:rho1}\\
\dot x_2 &=[p(0)+(p_u(0)+r(0))x_1^2+(p_u(0)-r(0))x_2^2+\cdots]x_2;
\label{eq:rho2}
\eeq
where $p(0)$ denotes the function $p(u,\Delta)$ evaluated at the origin. From
the linear terms in (\ref{eq:bhopf}) we find $p(0)={\rm Re}\lm$ and the third
order coefficients $p_u(0)$ and $r(0)$ are evaluated in the following section.

Assuming the non-degeneracy condition, $r(0)\,p(0)\,(p_u(0)-r(0))\neq0,$
then the amplitude equations (\ref{eq:rho1}) - (\ref{eq:rho2}) may be truncated
at third order.\cite{gss} In addition to the incoherent state
$(x_1,x_2)=(0,0)$, setting $\dot x_1=\dot x_2=0$ yields two other types of
solutions summarized in Table 1. There are travelling wave (TW) solutions
$(x_{{\rm TW}},0)$ and $(0,x_{{\rm TW}})$ with ${\rm Re}\lm
+(p_u(0)-r(0))\,x_{{\rm TW}}^2=0$ and standing wave solutions (SW) $(x_{{\rm
SW}},x_{{\rm SW}})$ with ${\rm Re}\lm +2p_u(0)\,x_{{\rm SW}}^2=0$.

In the rotating frame, both TW solutions have same frequency $\om_{{\rm
TW}}=q+x_{{\rm TW}}^2 s$ but the two states propagate oppositely in phase; the
$(x_{{\rm TW}},0)$ state corresponds to a density depending on
$(\theta+\om_{{\rm TW}}\,t)$ and the $(0,x_{{\rm TW}})$ state evolves according
to $(\theta-\om_{{\rm TW}}\,t)$. Note that $q$ and $s$ are evaluated on the TW
solution, and therefore near the bifurcation $\om_{{\rm TW}}$ is fixed by the
eigenvalue (\ref{eq:hev}) at criticality ($K=K_h)$:
\beq
\om_{{\rm TW}}^c&=&q(0)+\;\mbox{\rm higher order terms in}\;x_{{\rm
TW}}^2\nonumber\\
&\approx&\sqrt{\omz^2-(\eps+D)^2}.\label{eq:critfreq}
\eeq
In the lab frame (\ref{eq:lab}), the frequencies of these two states are split
to $\overline{\om}-\om_{{\rm TW}}$ and $\overline{\om} +\om_{{\rm TW}}$,
respectively.

The SW frequency is $\om_{{\rm SW}}=q$ where $q$ is evaluated on the SW
solution; near criticality $\om_{{\rm SW}}^c$ is also given by
(\ref{eq:critfreq}). The physical appearance of the SW can be described from
its form near onset,
\beq
\rho_{{\rm SW}}(\theta,\om,t)&=&\frac{1}{2\pi}+K\,x_{{\rm SW}}\mbox{Im}
\left[\frac{\alpha_0}{\om+\om_{{\rm SW}}^c-iD}\,e^{i\,(\theta+\om_{{\rm
SW}}^ct)}-\frac{\beta_0}{\om-\om_{{\rm SW}}^c+iD}\,e^{-i\,(\theta-\om_{{\rm
SW}}^ct)}\right]\nonumber\\
&&+\;\mbox{\rm higher order terms in}\;x_{{\rm SW}}\label{eq:critsw}
\eeq
where $\alpha_0$ and $\beta_0$ are initial phases.
There are two counter-propagating synchronized clumps containing oscillators
with native frequencies near $-\om_{{\rm SW}}^c$ and $+\om_{{\rm SW}}^c$,
respectively. Thus the population exhibits two simultaneous macroscopic
oscillations. In the lab frame these macroscopic oscillators appear as a slow
oscillator at frequency $\overline{\om}-\om_{{\rm SW}}$ and a fast oscillator
at $\overline{\om}+\om_{{\rm SW}}$.

The direction of bifurcation for these waves is determined by $(p_u(0)-r(0))$
and $p_u(0)$. Their linear stabilities are easily calculated: the TW
eigenvalues are $2(p_u(0)-r(0))\,x_{{\rm TW}}^2$ and ${\rm
Re}\lm+(p_u(0)+r(0))\,x_{{\rm TW}}^2$; the SW eigenvalues are $4p_u(0)\,x_{{\rm
SW}}^2$ and $-4r(0)\,x_{{\rm SW}}^2$. Depending on the signs and relative
magnitudes of $p_u(0)$ and $r(0)$ there are six possible bifurcations as shown
in Fig. 2. For the bimodal population (\ref{eq:bimode}) we shall see that
$(p_u(0)-r(0))$ and $p_u(0)$ are negative while $r(0)$ is positive. Thus the SW
are supercritical and stable and the TW are supercritical and unstable.
\begin{table}
\begin{center}
Table 1.  Branching equations and eigenvalues for non-degenerate Hopf
bifurcation
\end{center}
\vspace{7mm}
\begin{center}
\begin{tabular}{ccc}
\underline{Solution type $(x_1,x_2)$}    &
\hspace{0.25in}\underline{Branching Equations}    &
\hspace{0.25in}\underline{Eigenvalues}\\
\\
Travelling wave (TW)&${\rm Re}\lambda +(p_u-r)\,x_{{\rm TW}}^2=0$ &
$2(p_u-r)\,x_{{\rm TW}}^2$, \\
$(x_{{\rm TW}},0)$ and $(0,x_{{\rm TW}})$& & ${\rm Re}\lm+(p_u+r)\,x_{{\rm
TW}}^2$\\
\\
Standing wave (SW)&${\rm Re}\lm +2p_u\,x_{{\rm SW}}^2=0$ & $4p_u\,x_{{\rm
SW}}^2$,\\
$(x_{{\rm SW}},x_{{\rm SW}})$& & $-4r\,x_{{\rm SW}}^2$
\end{tabular}
\end{center}
\par
\vspace{7mm}
\underline{NOTES:}
\par
The coefficients $p_u$ and $r$ are evaluated at $(x_1,x_2)=(0,0)$.
\protect\vspace*{\fill}

\end{table}

\subsubsection{Evaluation of $p_u$ and $r$}

\move From (\ref{eq:h2hopf}) for $H_2$,
\beq
<\aeff,H_2>&=&\alpha^2\,<\aeff,h_2^{(2)}> +\alpha\beta^\ast\,<\aeff,h_2^{(1)}>+
(\beta^\ast)^2\,<\aeff,h_2^{(0)}> +\ord 3\\
<\k\cdot\aeff,H_{-2}>&=&(\alpha^\ast)^2\,<\k\cdot\aeff,{h_2^{(2)\ast}}>
+\alpha^\ast\beta\,<\k\cdot\aeff,{h_2^{(1)\ast}}>+ \nonumber\\
&&\hspace{1.0in}\beta^2\,<\k\cdot\aeff,{h_2^{(0)\ast}}> +\ord 3
\eeq
which determines the nonlinear terms in (\ref{eq:ahopf})-(\ref{eq:bhopf})
through third order:
\beq
\lefteqn{\left(\begin{array}{c}\dot{\alpha}\\ \dot{\beta}\end{array}\right)=}
\nonumber\\
&&\lm\left(\begin{array}{c}\alpha\\ \beta\end{array}\right)
-\pi K \left(\begin{array}{c}
\alpha[<\aeff,h_2^{(2)}>|\alpha|^2+<\aeff,h_2^{(1)}>|\beta|^2]\\
\beta[<\k\cdot\aeff,{h_2^{(0)\ast}}>|\beta|^2 +
<\k\cdot\aeff,{h_2^{(1)\ast}}>|\alpha|^2]
\end{array}\right)\nonumber\\
&&\nonumber\\
&&\hspace{0.5in}-\pi K \left(\begin{array}{c}
\alpha^\ast[\alpha\beta^\ast\,<\aeff,h_2^{(1)}>+
(\beta^\ast)^2\,<\aeff,h_2^{(0)}>]\\
\beta^\ast[(\alpha^\ast)^2\,<\k\cdot\aeff,{h_2^{(2)\ast}}>
+\alpha^\ast\beta\,<\k\cdot\aeff,{h_2^{(1)\ast}}>]
\end{array}\right)\nonumber\\
&&\nonumber\\
&&\hspace{0.5in}-\pi K \left(\begin{array}{c}
\beta [\alpha^2\,<\aeff,h_2^{(2)}> + (\beta^\ast)^2\,<\aeff,h_2^{(0)}>]\\
\alpha [(\alpha^\ast)^2\,<\k\cdot\aeff,{h_2^{(2)\ast}}> +
\beta^2\,<\k\cdot\aeff,{h_2^{(0)\ast}}>]
\end{array}\right)+\ord 4\label{eq:hopfcm3}
\eeq

Only the first set of  nonlinear terms in (\ref{eq:hopfcm3}) have the phase
shift symmetry and are retained in the normal form:
\beq
\left(\begin{array}{c}\dot{\alpha'}\\ \dot{\beta'}\end{array}\right)&=&
\lm\left(\begin{array}{c}\alpha'\\ \beta'\end{array}\right)
-\pi K \left(\begin{array}{c}
\alpha'[<\aeff,h_2^{(2)}>|\alpha'|^2+<\aeff,h_2^{(1)}>|\beta'|^2]\\
\beta'[<\k\cdot\aeff,{h_2^{(0)\ast}}>|\beta'|^2 +
<\k\cdot\aeff,{h_2^{(1)\ast}}>|\alpha'|^2]
\end{array}\right)\label{eq:hopf3nf}\\
&&\hspace{1.0in}+\ord 5;\nonumber
\eeq
the remaining cubic terms and all fourth order terms in (\ref{eq:hopfcm3}) are
removed by the normal form change of variables
$(\alpha,\beta)\rightarrow(\alpha',\beta')$.
{}From (\ref{eq:h0id}) and (\ref{eq:h1id}) the coefficients in
(\ref{eq:hopf3nf}) are related:
$<\aeff,h_2^{(2)}>=<\k\cdot\aeff,{h_2^{(0)\ast}}>$ and
$<\aeff,h_2^{(1)}>=<\k\cdot\aeff,{h_2^{(1)\ast}}>$ so that the result in
(\ref{eq:hopf3nf}) can be rewritten in the standard form (\ref{eq:hopfnf0}).
Noting that $|\alpha'|^2={(u-\delta)}/{2}$ and $|\beta'|^2={(u+\delta)}/{2}$,
(\ref{eq:hopf3nf}) becomes
\beq
\left(\begin{array}{c}\dot{\alpha'}\\ \dot{\beta'}\end{array}\right)&=&
\left[\lm-\frac{\pi K}{2} (<\aeff,h_2^{(2)}>+<\aeff,h_2^{(1)}>)\,u+\ord
4\right] \left(\begin{array}{c}\alpha'\\ \beta'\end{array}\right)\nonumber\\
&&+\left[\frac{\pi K}{2}(<\aeff,h_2^{(2)}>-<\aeff,h_2^{(1)}>)+\ord
2\right]\,\delta\left(\begin{array}{c}\alpha'\\ -\beta'\end{array}\right).
\label{eq:stdnf}
\eeq
By comparing (\ref{eq:hopfnf0}) and (\ref{eq:stdnf}) we can identify the cubic
coefficients $p_u$ and $r$
\beq
p_u(0)&=&-\frac{\pi K}{2}\,{\rm Re}\,<\aeff,[h_2^{(1)}+h_2^{(2)}]>
\label{eq:pu}\\
r(0)&=&-\frac{\pi K}{2}\,{\rm Re}\,<\aeff,[h_2^{(1)}-h_2^{(2)}]>\label{eq:r}
\eeq
in terms of the center manifold coefficients $h_2^{(2)}$ and $h_2^{(1)}$.

For this bifurcation the equation for $H$ (\ref{eq:Heqn}) becomes
\beq
\lefteqn{\left[\frac{\partial H}{\partial \alpha}\dot{\alpha} + \frac{\partial
H}{\partial \beta}\dot{\beta}+cc\right]=}\label{eq:hopfcm}\\
&&\lop H -\pi K
\left\{ e^{i\th} (\alpha^\ast +\beta +<g,H_{-1}>)
[H_2-<\aeff,H_2>\eff-<\k\cdot\aeff^\ast,H_2>\k\cdot\eff^\ast]\right.\nonumber\\
&&+2\,e^{i2\th} \left[(\alpha^\ast +\beta+ <g,H_{-1}>)H_{3}(\om)-
(\alpha+\beta^\ast+ <g,H_{1}>)(\alpha\,\eff+ \beta^\ast\,\k\cdot\eff^\ast
+H_1)\right]\nonumber\\
&&+\sum_{l\geq3}\,l\,e^{il\th} \left[
(\alpha^\ast + \beta+ <g,H_{-1}>)H_{l+1}(\om) -
\left. (\alpha+ \beta^\ast+ <g,H_{1}>)H_{l-1}(\om)\right]+cc\right\},\nonumber
\eeq
and the functions $h_2^{(2)}$, $h_2^{(1)}$, and $h_2^{(0)}$ may be determined
from (\ref{eq:hopfcm}). By inspection of the $l=2$ component we note that the
inhomogeneous terms $2\pi\,K\,e^{i2\th}\,(\alpha+\beta^\ast)\,(\alpha\,\eff+
\beta^\ast\,\k\cdot\eff^\ast)$ in (\ref{eq:hopfcm}) generate the leading terms
$h_2^{(i)}(\om)$ in $H_2$; since $H_l(\om)\sim\ord 3$ for $l\neq2$, applying
(\ref{eq:h2hopf}) to the $l=2$ component of (\ref{eq:hopfcm}) and equating
terms of the same order, we obtain
\beq
2\lm\,h_2^{(2)}&=&\flop 2 h_2^{(2)} + 2\pi\,K\,\eff\\
(\lm+\lm^\ast)\,h_2^{(1)}&=&\flop 2 h_2^{(1)} +
2\pi\,K\,(\eff+\k\cdot\eff^\ast)\\
2\lm^\ast\,h_2^{(0)}&=&\flop 2 h_2^{(0)}+ 2\pi\,K\,\k\cdot\eff^\ast,
\eeq
with solutions
\beq
h_2^{(2)}(\om)&=&\frac{-i\pi\,K\,\eff(\om)}{\om-z_0-i2D}\label{eq:h1soln}\\
h_2^{(1)}(\om)&=&\frac{-i\pi\,K\,(\eff(\om)+\k\cdot\eff(\om)^\ast)}{\om-i\,{\rm
Im}\,z_0-i2D}
\label{eq:h2soln}\\
h_2^{(0)}(\om)&=&\frac{-i\pi\,K\,\k\cdot\eff(\om)^\ast}{\om+z_0^\ast-i2D}.
\label{eq:h3soln}
\eeq
These results satisfy the relations (\ref{eq:h0id}) - (\ref{eq:h1id}).

{}From (\ref{eq:pu}) - (\ref{eq:r}) these results determine the cubic
coefficients explicitly
\beq
p_u(0)&=&\frac{-\pi^2 K^2}{2}\,{\rm Im}\,
\int_{-\infty}^{\infty}\,d\om\,\aeff(\om)^\ast\,
\left[\frac{\eff(\om)+\k\cdot\eff(\om)^\ast}{(\om-i{\rm Im}\,z_0-i2D)}
+\frac{\eff(\om)}{(\om-z_0-i2D)}\right]\\
&&\nonumber\\
r(0)&=&\frac{-\pi^2 K^2}{2}\,{\rm Im}\,
\int_{-\infty}^{\infty}\,d\om\,\aeff(\om)^\ast\,
\left[\frac{\eff(\om)+\k\cdot\eff(\om)^\ast}{(\om-i{\rm Im}\,z_0-i2D)}
-\frac{\eff(\om)}{(\om-z_0-i2D)}\right].
\eeq
Provided Im $z_0>-D$,  any singularities in the lower half plane are due to
$g(\om)$. Hence by inspection we can see that at criticality for the
bifurcation (Im $z_0=0$), the limit $D\rightarrow0$ is finite for both $p_u(0)$
and $r(0)$. By contrast the corresponding limit for the Vlasov equation
involves pinching singularities that produce a singular $D^{-3}$
behavior.\cite{jdc} Here the well-behaved limit at $D=0$ suggests that the
oscillations produced by the bifurcation will grow like $\sqrt{|K-K_c|}$ even
in the absence of noise.

\subsubsection{Example of a bimodal population}

\move We illustrate these results for the bimodal population described in
(\ref{eq:bimode}); for $4\omz>K$ the eigenvalues $(\lambda,\lambda^\ast)$ are
given by $\lambda=-iz_0=\,(K-K_h)/4\pm i\,\Omega$
where $\Omega\equiv\sqrt{\omz^2-(K/4)^2}$ and $K_h\equiv 4(\eps+D)$. If $K$
increases above $K_h$, there is a Hopf bifurcation.
The coefficients $p_u(0)$ and $r(0)$ can be evaluated analytically at the
bifurcation $K=K_h$; by closing the integration contours in the lower half
plane and using $\Omega^2=\omz^2-(\eps+D)^2$ one obtains
\beq
\sfcnp{1}{z_0}&=&\frac{iK_h}{2}\,\frac{\omz^2+(\Omega-i(\eps+D))^2}
{[\omz^2-(\Omega-i(\eps+D))^2]^2}\\
&&\nonumber\\
{\rm Re}\,<\aeff,h_2^{(1)}>&=&\frac{16\pi K_h D\,(\eps+D)^2\Omega^2\omz^4}
{|\omz+i(\eps+2D)|^2\;|\omz^4-(\Omega+i(\eps+D))^4\;|^2}\label{eq:h1}\\
&&\nonumber\\
{\rm Re}\,<\aeff,h_2^{(2)}>&=&
\frac{4\pi K_h\Omega^2\; (\eps+2D)\,\omz^2\,[4\omz^2+(\eps+2D))^2-(\eps+D))^2]}
{|\omz^2+(\Omega-i(\eps+D))^2|^2\;|\omz^2-(\Omega-i(\eps+2D))^2|^2}.
\label{eq:h2}
\eeq

The direction of bifurcation and stability of the travelling waves and standing
waves are controlled by $p_u(0)$, $r(0)$, and $p_u(0)-r(0)$. Since ${\rm
Re}\,<\aeff,h_2^{(1)}>$ and ${\rm Re}\,<\aeff,h_2^{(2)}>$ are clearly positive,
one can see by inspection that both $p_u(0)$ and $p_u(0)-r(0)$ are negative,
and the SW and TW solutions are always supercritical. The bifurcation produces
stable TW solutions if $r(0)<0$ and stable SW if $r(0)>0$; see Fig. 2.
Subtracting ${\rm Re}\,<\aeff,h_2^{(2)}>$ from ${\rm Re}\,<\aeff,h_2^{(1)}>$
gives
\be
r(0)=\frac{8\pi^2 K_h^2\Omega^2\omz^4\,(\eps+D)^2\;P(\eps,D,\omz^2)}
{|\omz^4-(\Omega+i(\eps+D))^4\;|^2\;|\omz^2-(\Omega-i(\eps+2D))^2|^2\;
|\omz+i(\eps+2D)|^2}\label{eq:r0}
\ee
where
\beq
P(\eps,D,\omz^2)&=& 4(\eps+2D)\,\omz^4+(\eps+2D)\,[4\eps^2+14\eps
D+11D^2]\,\omz^2\nonumber\\
&&+D(\eps+D)\,(3\eps+3D)\,[\eps^2+5\eps D+5 D^2].\label{eq:P}
\eeq
Thus for this bimodal population $r(0)$ is always positive and the bifurcation
always leads to stable standing waves.

In light of Kuramoto's earlier discussion of the partially synchronized bimodal
population (cf. the remarks in Section 2.3), it is interesting to compare the
frequencies $\om_{\rm peak}$ in (\ref{eq:peaks}) and $\om_{{\rm SW}}^c$ in
(\ref{eq:critfreq}) for the SW near onset in the relevant frequency range
$(1+D/\eps)<\omz/\eps<\infty$. From (\ref{eq:peaks}) and (\ref{eq:critfreq}) we
find
\be
\left(\frac{\om_{{\rm SW}}^c}{\om_{\rm peak}}\right)^2=
\frac{(\omz/\eps)^2-(1+D/\eps)^2}
{\left[(2\omz/\eps)-\sqrt{1+(\omz/\eps)^2}\right]\sqrt{1+(\omz/\eps)^2}};\label{eq:ratio}
\ee
as the peaks move far apart ($\omz/\eps\rightarrow\infty$) this ratio
approaches unity and the oscillators in the peaks are indeed the oscillators
that synchronize. However as $\omz/\eps$ approaches $(1+D/\eps)$ the
oscillators comprising the standing wave synchronize at frequencies that are
substantially shifted away from the peaks of the native distribution towards
the interior of the distribution; this is illustrated in Fig. 3.

\subsubsection{Comparison to Bonilla, Neu, and Spigler}

\move Bonilla, {\em et al.} have previously calculated the amplitude equation
for a Hopf bifurcation from the incoherent state, but they neglected to include
the second eigenvector $\k\cdot\ef$ in (\ref{eq:hopfev2}) that is forced by
$\otwo$ symmetry. In our notation this omission has the effect of setting the
amplitude $\beta(t)$ of this mode to zero; if this is done then the
two-dimensional system for $(x_1,x_2)$ in (\ref{eq:rho1}) - (\ref{eq:rho2}) is
reduced to\footnote{This corresponds to equation (3.18a) in Bonilla, {\em et
al.} with the identification $Kx_1/2\rightarrow |A|$.}
\be
\dot x_1 =[p(0)+(p_u(0)-r(0))x_1^2+\cdots]x_1.\label{eq:bns}
\ee
Since $p_u(0)-r(0)<0$ for the bimodal population, this equation predicts a
single TW $(x_1,x_2)=(x_{{\rm TW}},0)$ which is supercritical and stable. The
SW solutions $(x_1,x_2)=(x_{{\rm SW}},x_{{\rm SW}})$ and the second TW solution
$(x_1,x_2)=(0,x_{{\rm TW}})$ are omitted; in addition the actual instability of
the TW is missed because the TW eigenvalue ${\rm Re}\lm+(p_u(0)+r(0))\,x_{{\rm
TW}}^2$ corresponding to perturbations in the direction of the SW state is not
included.

\subsubsection{Comparison to Okuda and Kuramoto}

\move Okuda and Kuramoto consider two populations of oscillators $(i=1,2)$ with
number densities $n_i(\phi,t)$; in the absence of couplings each population has
a definite native frequency: $\om_1=\overline{\om}+\Delta\om/2$ and
$\om_2=\overline{\om}-\Delta\om/2$ respectively. The evolution of $n_1(\phi,t)$
and $n_2(\phi,t)$ is described by
\beq
\frac{\partial n_1}{\partial t}&=& -\om_1\frac{\partial n_1}{\partial \phi}
+\frac{\partial }{\partial \phi}\left\{n_1(\phi,t)
\int_0^{2\pi}\,d\phi'\,\sin(\phi-\phi')\,[K_1'n_1(\phi',t)+
K'n_2(\phi',t)]\right\}\nonumber\\
&&+D\frac{\partial^2 n_1}{\partial\phi^2 }\label{eq:n1}\\
\frac{\partial n_2}{\partial t}&=&-\om_2\frac{\partial n_2}{\partial \phi}
+\frac{\partial }{\partial \phi}\left\{n_2(\phi,t)
\int_0^{2\pi}\,d\phi'\,\sin(\phi-\phi')\,[K_2'n_2(\phi',t)+
K'n_1(\phi',t)]\right\}\nonumber\\
&&+D\frac{\partial^2 n_2}{\partial\phi^2 }.\label{eq:n2}
\eeq
The intra-population couplings are $K_1'$ and $K_2'$ and the coupling between
the populations is $K'$. The incoherent state is an equilibrium described by
$n_1(\phi)=n_2(\phi)=1/2\pi$, and the authors analyzed bifurcations from this
state. This model has $\otwo$ symmetry if $\overline{\om}=0$ and $K_1'=K_2'$.

For the bimodal population (\ref{eq:bimode}) at $\eps=0$, the Fokker-Planck
equation (\ref{eq:eveqn}) reduces to the O-K model but with the constraint of
equal couplings $K_1'=K_2'=K'$. For the $\eps=0$ distribution,
\be
g(\om)=\frac{1}{2}\left[\delta(\om+\omz)+\delta(\om-\omz)\right],
\ee
we define number densities by integrating over the frequency distribution of
each sub-population
\beq
n_1(\theta,t)&=&\int_{-\infty}^{\infty}\,d\om\,\delta(\om-\omz)\,\rho(\theta,\omega,t)\\
&&\\
n_2(\theta,t)&=&\int_{-\infty}^{\infty}\,d\om\,\delta(\om+\omz)\,\rho(\theta,\omega,t)
\eeq
so that (\ref{eq:eveqn}) becomes
\beq
\frac{\partial n_1}{\partial t}&=& -\omz\frac{\partial n_1}{\partial \theta}
+\frac{\partial }{\partial \theta}\left\{n_1(\theta,t)
\int_0^{2\pi}\,d\theta'\,\sin(\theta-\theta')\,[(K/2)\,n_1(\theta',t)+
(K/2)\,n_2(\theta',t)]\right\}\nonumber\\
&&+D\frac{\partial^2 n_1}{\partial\theta^2 }\label{eq:n1fp}\\
\frac{\partial n_2}{\partial t}&=&+\omz\frac{\partial n_2}{\partial \theta}
+\frac{\partial }{\partial \theta}\left\{n_2(\theta,t)
\int_0^{2\pi}\,d\theta'\,\sin(\theta-\theta')\,[(K/2)\, n_2(\theta',t)+
(K/2)\,n_1(\theta',t)]\right\}\nonumber\\
&&+D\frac{\partial^2 n_2}{\partial\theta^2 }.\label{eq:n2fp}
\eeq
The equations (\ref{eq:n1}) - (\ref{eq:n2}) have this form if we set
$\Delta\om=2\omz$ and $K_1'=K_2'=K'=K/2$ and transform to the rotating frame
$\theta=\phi-\overline{\om}\,t$ where average frequency is zero. One of the
cases considered by Okuda and Kuramoto is $K_1'=K_2'=1$ with $K'$ and
$\Delta\om$ variable. Their results for the steady state bifurcation and Hopf
bifurcation are consistent with those of this paper. In particular, for the
Hopf bifurcation they find that the SW solutions are expected to be stable.

\subsubsection{Perturbing $\otwo\rightarrow\sotwo$}

\move If the distribution $g(\om)$ is perturbed so that the reflection symmetry
(\ref{eq:g}) is broken, then the complex eigenvalues of multiplicity-two split
into two non-degenerate complex conjugate pairs. The Hopf bifurcation
considered here perturbs to two distinct Hopf bifurcations corresponding to the
distinct pairs of eigenvalues.  The resulting Hopf bifurcations individually
lead to time-periodic states in the form of rotating or travelling waves and
the interaction between these waves yields modulated waves via secondary
bifurcations.\cite{baj}-\cite{ph} The modulated waves are the perturbed form of
the standing waves discussed above; Knobloch discusses such perturbed
bifurcations in more detail.\cite{kno}

\section{\hspace{0.125in}Discussion}
\move The bifurcation analysis is summarized in Fig. 4 which shows the stable
nonlinear states that arise from instabilities of the incoherent state. Our
discussion does not treat the transition from the stable partially synchronized
state to the stable standing waves; this transition is represented by a dashed
boundary in Fig. 4. The interaction of the standing waves and the synchronized
state can be studied by analyzing the codimension-two bifurcation defined
by the merger of the Hopf and steady state bifurcations. The linear theory of
the non-semisimple real eigenvalue was developed in Section 2 of this paper and
a normal form theory for the $\otwo$ Takens-Bogdanov bifurcation has been given
by Dangelmayr and Knobloch\cite{dk}. One complicating feature of the bimodal
population is the subcriticality of the steady state bifurcation at the
codimension-two point; one would like to include the stable partially
synchronized state at finite amplitude in the analysis. This could be done by
generalizing the Dangelmayr-Knobloch theory to combine the nonlinear degeneracy
of the steady state bifurcation with the linear degeneracy. Formally this leads
to a codimension-three bifurcation and would require a normal form of higher
order; the analysis of Dangelmayr and Knobloch truncates the normal form
equations after cubic terms.

\section{\hspace{0.125in}Acknowledgements}

I have enjoyed helpful conversations with Steve Strogatz.

\section{\hspace{0.125in}Figure captions}
\begin{enumerate}

\item Eigenvalues of the incoherent state as a function of $\eps$,$D$, $K$, and
$\omz$. The solid curves indicate parameter values at which there is an
eigenvalue on the imaginary axis; along $K=K_c$ there is an eigenvalue at zero
and along $K=K_h$ there is a pure imaginary complex-conjugate pair. At the
juncture $K=4(\eps+D)$ and $\omz=(\eps+D)$, is the codimension two point where
the Hopf bifurcation surface intersects the steady state bifurcation surface.
The dashed curves indicate parameter values where eigenvalues coalesce with the
continuum; upon crossing $K=K_e'$ or $K=K_e$ an eigenvalue either appears or
disappears. Along the dot-dashed line $K=4\omz$, the two eigenvalues collide on
the real axis. The inset diagrams show the qualitative features of the
eigenvalue spectrum in each region. The continuous spectrum at $-D$ is not
drawn.

\item Bifurcation diagrams for Hopf bifurcation with $\otwo$ symmetry. The
stability and direction of bifurcation for the travelling waves (TW) and
standing waves (SW) are determined by the two cubic coefficients $r(0)$ and
$p_u(0)$. The incoherent state is the horizontal branch and the two TW
solutions are drawn as a single branch. Stable solutions are solid branches and
unstable solutions are dashed. The signs of the real parts of the eigenvalues
are indicated; see Table 1 for the TW and SW eigenvalues.

\item Comparison of the standing wave frequency $\om_{{\rm SW}}^c$ and
$\om_{\rm peak}$ for the bimodal distribution  at criticality $K=K_h$ for the
Hopf bifurcation along the line $(1+D/\eps)<(\omz/\eps)<\infty$. Also shown is
the ratio of $\omz$ to $\om_{\rm peak}$; the curves are drawn for $D/\eps=1$.

\item Nonlinear states for the bimodal distribution: the incoherent state (I),
the partially synchronized state (PS), and the standing waves (SW). The solid
lines are the stability boundaries for steady state bifurcation and Hopf
bifurcation. The boundary between the standing waves and the partially
synchronized state is shown schematically as a dashed line; the precise nature
and location of this boundary has not been determined.

\end{enumerate}

\clearpage

\end{document}